\documentstyle[psfig]{mn}
\newif\ifAMStwofonts

\newcommand{\delc}{\delta_{\rm c}}
\newcommand{\sgw}{\sigma^2}
\newcommand{\msun}{M_{\odot}}
\newcommand{\abs}[1]{\left|#1\right|}
\newcommand{\lbkt}[1]{\left[#1\right]}
\newcommand{\mbkt}[1]{\left\{#1\right\}}
\newcommand{\sbkt}[1]{\left(#1\right)}

\newcommand{\ngt}{\raisebox{-0.8ex}{~$\stackrel{\textstyle >}{\sim}$}~}
\newcommand{\nlt}{\raisebox{-0.8ex}{~$\stackrel{\textstyle <}{\sim}$}~}
\ifoldfss
  \ifCUPmtlplainloaded \else
    \NewTextAlphabet{textbfit} {cmbxti10} {}
    \NewTextAlphabet{textbfss} {cmssbx10} {}
    \NewMathAlphabet{mathbfit} {cmbxti10} {} % for math mode
    \NewMathAlphabet{mathbfss} {cmssbx10} {} %  "   "    "
  \fi
  \ifAMStwofonts
    \ifCUPmtlplainloaded \else
      \NewSymbolFont{upmath} {eurm10}
      \NewSymbolFont{AMSa} {msam10}
      \NewMathSymbol{\upi}     {0}{upmath}{19}
      \NewMathSymbol{\umu}     {0}{upmath}{16}
      \NewMathSymbol{\upartial}{0}{upmath}{40}
      \NewMathSymbol{\leqslant}{3}{AMSa}{36}
      \NewMathSymbol{\geqslant}{3}{AMSa}{3E}

    \fi
  \fi
\fi % End of OFSS

\ifnfssone
  \newmathalphabet{\mathit}
  \addtoversion{normal}{\mathit}{cmr}{m}{it}
  \addtoversion{bold}{\mathit}{cmr}{bx}{it}
  \newmathalphabet{\mathbfit} % math mode version of \textbfit{..}
  \addtoversion{normal}{\mathbfit}{cmr}{bx}{it}
  \addtoversion{bold}{\mathbfit}{cmr}{bx}{it}
  \newmathalphabet{\mathbfss} % math mode version of \textbfss{..}
  \addtoversion{normal}{\mathbfss}{cmss}{bx}{n}
  \addtoversion{bold}{\mathbfss}{cmss}{bx}{n}
  \ifAMStwofonts
    \ifCUPmtlplainloaded \else
      \UseAMStwoboldmath
      \makeatletter
      \new@mathgroup\upmath@group
      \define@mathgroup\mv@normal\upmath@group{eur}{m}{n}
      \define@mathgroup\mv@bold\upmath@group{eur}{b}{n}
      \edef\UPM{\hexnumber\upmath@group}
      \new@mathgroup\amsa@group
      \define@mathgroup\mv@normal\amsa@group{msa}{m}{n}
      \define@mathgroup\mv@bold\amsa@group{msa}{m}{n}
      \edef\AMSa{\hexnumber\amsa@group}
      \makeatother
      \mathchardef\upi="0\UPM19
      \mathchardef\umu="0\UPM16
      \mathchardef\upartial="0\UPM40
      \mathchardef\leqslant="3\AMSa36
      \mathchardef\geqslant="3\AMSa3E
    \fi
  \fi
\fi % End of NFSS release 1

\ifnfsstwo
  \DeclareMathAlphabet{\mathbfit}{OT1}{cmr}{bx}{it}
  \SetMathAlphabet\mathbfit{bold}{OT1}{cmr}{bx}{it}
  \DeclareMathAlphabet{\mathbfss}{OT1}{cmss}{bx}{n}
  \SetMathAlphabet\mathbfss{bold}{OT1}{cmss}{bx}{n}
  \ifAMStwofonts
    \ifCUPmtlplainloaded \else
      \DeclareSymbolFont{UPM}{U}{eur}{m}{n}
      \SetSymbolFont{UPM}{bold}{U}{eur}{b}{n}
      \DeclareSymbolFont{AMSa}{U}{msa}{m}{n}
      \DeclareMathSymbol{\upi}{0}{UPM}{"19}
      \DeclareMathSymbol{\umu}{0}{UPM}{"16}
      \DeclareMathSymbol{\upartial}{0}{UPM}{"40}
      \DeclareMathSymbol{\leqslant}{3}{AMSa}{"36}
      \DeclareMathSymbol{\geqslant}{3}{AMSa}{"3E}
    \fi
  \fi
\fi % End of NFSS release 2

\ifCUPmtlplainloaded \else
  \ifAMStwofonts \else % If no AMS fonts
    \def\upi{\pi}
    \def\umu{\mu}
    \def\upartial{\partial}
  \fi
\fi

\title[Formation rate of gravitational structures]
  {\vspace{5mm} Formation rate of gravitational structures  
  and the cosmic X-ray background radiation}
\author[T. Kitayama  and  Y. Suto]
  {Tetsu Kitayama\thanks{e-mail:
kitayama@utaphp1.phys.s.u-tokyo.ac.jp} 
and Yasushi Suto\thanks{e-mail: suto@phys.s.u-tokyo.ac.jp, also
at Research Center for the Early Universe, School of Science,
The University of Tokyo, Bunkyo-ku, Tokyo 113, Japan 
} \\ 
 Department of Physics, 
The University of Tokyo,  
           Bunkyo-ku, Tokyo 113, Japan 
}
\date{Accepted 1995 November 27}
\pagerange{\pageref{firstpage}--\pageref{lastpage}}
\pubyear{1996}

\def\LaTeX{L\kern-.36em\raise.3ex\hbox{a}\kern-.15em
    T\kern-.1667em\lower.7ex\hbox{E}\kern-.125emX}

\begin{document}

\label{firstpage}

\maketitle

\begin{abstract}
 Analytical expressions for the rates of formation and destruction of
gravitationally bound systems are derived assuming that they are
originated from primordial random-Gaussian density fluctuations.  The
resulting formulae reproduce the time derivative of the
Press-Schechter mass function in a certain limit.  Combining a
theoretical model for the evolution of structures with the formation
rate, we can make various cosmological predictions which are to be
compared with observations.  As an example to elucidate such
applicability, we evaluate the contribution of clusters of galaxies to
the cosmic X-ray background radiation. With the {\it COBE}
normalization, we find that the significant fraction of the observed
soft X-ray background is accounted for by clusters of galaxies in a
cold dark matter universe with $\Omega_0 \sim 0.2$, $\lambda_0 =1-
\Omega_0$ and $h \sim 0.8$.
\end{abstract}


\section{Introduction}

Formation, merging, and destruction history of dark matter haloes of
astrophysical objects is of fundamental importance in describing the
subsequent formation and evolution of luminous objects such as
quasars, galaxies, and clusters of galaxies. In addition, early
formation of structures may affect the growth of objects through the
reionization of the entire universe (Sasaki, Takahara, \& Suto 1993;
Sasaki \& Takahara 1994; Fukugita \& Kawasaki 1994).

Key quantities in this context are the rates of formation and
destruction of gravitationally bound systems. They are defined as the
comoving number density of bound systems of a given mass that are
formed or destroyed in unit time at a given epoch.  A number of
authors have attempted to compute this quantity either numerically or
analytically. A straightforward approach is to use a state-of-art
numerical simulations (e.g., Evrard 1990; Cen \& Ostriker 1992; Bryan
et al. 1994; Suginohara 1994). Although the dynamical range and
small-scale resolution of available codes are rapidly increasing, it
is not clear at present to what extent they unambiguously describe
small-scale dynamics. Besides it is not easy to grasp the underlying
key physics from the results of numerical simulations. It is therefore
important to develop an analytical model on the basis of simple
physics.

The alternative approach in such a direction is closely related to the
theory of Press \& Schechter (1974; hereafter PS). PS derived a mass
function, the number density as a function of mass and epoch, of
virialized systems.  The PS mass function, however, is not directly
related to the rate of formation mentioned above.  In fact, the time
derivative of the PS mass function can become either positive or
negative. Thus it can not be simply interpreted as the rate of
formation, but rather corresponds to the balance between the rates of
formation and destruction (Cavaliere, Colafrancesco \& Scaramella
1991; Blain \& Longair 1993a,b).  Although several authors have
adopted the time derivative of the PS mass function to approximate the
formation rate of bound systems (e.g., Efstathiou \& Rees 1988;
Fukugita \& Kawasaki 1994), its validity should be examined carefully.
Blain \& Longair (1993a,b) and Sasaki (1994) correctly realized the
above problem, and attempted to obtain the rates of formation and
destruction separately, assuming some empirical forms for the rates.
In this paper, we derive a more general formula for each rate on the
basis of merger probabilities discussed by Bower (1991) and Lacey \&
Cole (1993; hereafter LC) (See also Kauffmann \& White 1993).

Our formalism is applicable to a number of problems regarding the
hierarchical clustering in the universe. In what follows we further
apply the resultant formulae to the formation of clusters of galaxies
and calculate their contribution to the observed cosmic X-ray
background (XRB). The cluster contribution is in general supposed to
be very small in the high energy band above a few keV, while it is
likely to dominate in the soft energy band (Silk \& Tarter 1973;
Schaeffer \& Silk 1988). It is, however, still uncertain to what
extent the observed XRB can be contributed by emissions from clusters
of galaxies. Several authors have employed the PS mass function to
evaluate the cluster contribution (Evrard \& Henry 1991; Blanchard et
al. 1992; Burg, Cavaliere \& Menci 1993), and thus formation epochs
and the subsequent evolution of virialized clusters are not well
specified in their treatment.  Here we present an approach which
explicitly takes account of the formation epoch of clusters and their
luminosity evolution.  Our predictions are made specifically in cold
dark matter (CDM) universes with/without the cosmological constant
whose fluctuation spectra are properly normalized by the {\it COBE}
data (Sugiyama 1995).  The results are also compared with those of the
hydrodynamical simulation and the {\it ASCA} observations.

The plan of this paper is as follows.  In Section 2, we outline the PS
formalism and discuss how the time derivative of its mass function is
related to the formation and destruction rates of bound systems. Then
using the merger probabilities of virialized haloes deduced by Bower
(1991) and LC, we derive expressions for the rates
of formation and destruction. These expressions are further employed
to describe the distribution of formation epoch of bound systems.  In
Section 3, we apply our formulae and calculate the contribution of
clusters of galaxies to the XRB.  Finally Section 4 is devoted to our
conclusions and further discussion.

\section{FORMATION AND DESTRUCTION RATES OF VIRIALIZED
          OBJECTS}
\subsection{Press-Schechter mass function}

According to the hierarchical clustering model, small density
fluctuations in the early universe grow via gravitational instability,
become nonlinear, and eventually collapse to form bound virialized
systems. For random Gaussian initial density fluctuations, the
comoving number density of spherical collapsed systems in the mass
range $M \sim M+dM$ at time $t$ is evaluated as (Press \& Schechter
1974; Bond et al. 1991)
%%%%%%%%%%%%%%%%%%%%%%%%%%%%%%%%%%%%%%%%%%%%%%%%%%%%%%%%%%%
\begin{eqnarray}
  \lefteqn{N_{\rm PS}(M,t) dM} \nonumber \\
 & &  = \sqrt{\frac{2}{\pi}} \frac{\rho_{0}}{M} 
    \frac{\delc(t)}{\sgw(M)} \abs{\frac{d\sigma(M)}{dM}} 
   \exp\lbkt{-\frac{\delc^{2}(t)}{2\sgw(M)}}dM ,  
\label{ps}
\end{eqnarray}
%%%%%%%%%%%%%%%%%%%%%%%%%%%%%%%%%%%%%%%%%%%%%%%%%%%%%%%%%%%
where $\rho_{0}$ is the mean comoving density of the universe,
$\delc(t)$ is the critical density threshold for a spherical
perturbation to collapse by the time $t$, and $\sigma(M)$ is the rms
density fluctuation smoothed over a region of mass $M$.  The latter
two quantities are expressed respectively as
%%%%%%%%%%%%%%%%%%%%%%%%%%%%%%%%%%%%%%%%%%%%%%%%%%%%%%%%%%%%%
\begin{equation}
  \delc(t)\equiv\frac{D(t_0)}{D(t)}\delc = \frac{\delc}{D(t)} 
\hspace{5mm} \mbox{($\delc\simeq 1.69$~ for~ $\Omega_0=1$)} ,
\label{deltc}
\end{equation}
%%%%%%%%%%%%%%%%%%%%%%%%%%%%%%%%%%%%%%%%%%%%%%%%%%%%%%%%%%%%%
and
%%%%%%%%%%%%%%%%%%%%%%%%%%%%%%%%%%%%%%%%%%%%%%%%%%%%%%%%%%%
\begin{equation}
 \sgw(M)\equiv \sigma^{2}(M,t_0) 
          = \frac{1}{(2\pi)^3}\int P(k,t_0) \hat{W}_{M}^{2}(k) 
             d^3 k, 
\label{variance}
\end{equation}
%%%%%%%%%%%%%%%%%%%%%%%%%%%%%%%%%%%%%%%%%%%%%%%%%%%%%%%%%%%
where $D(t)$ is the linear growth factor normalized to unity at the
present epoch $t_0$, $\Omega_0$ is the cosmological density parameter,
$P(k,t_0)$ is the present power spectrum of density fluctuations, and
$\hat{W}_{M}(k)$ is the Fourier transform of a real space window
function whose volume contains mass $M$.

Let us now consider the time derivative of the PS mass function
(\ref{ps}):
%%%%%%%%%%%%%%%%%%%%%%%%%%%%%%%%%%%%%%%%%%%%%%%%%%%%%%%%%%%%%%%
\begin{eqnarray}
  \lefteqn{ R_{\rm PS}(M,t) \equiv {d N_{\rm PS}(M,t) \over dt}} \nonumber \\
     & & = \lbkt{\frac{\delc(t)}{\sgw(M)}-\frac{1}{\delc(t)}}
                       \lbkt{-\frac{d\delc(t)}{dt}}N_{\rm PS}(M,t) .
\label{rps}
\end{eqnarray}
%%%%%%%%%%%%%%%%%%%%%%%%%%%%%%%%%%%%%%%%%%%%%%%%%%%%%%%%%%%%%%%
The above expression is positive only for $\sigma(M) < \delc(t)$ since
$d\delc(t)/dt<0$. This implies that $R_{\rm PS}(M,t)$ is negative for
low-mass objects for almost all realistic power spectra and window
functions. Consider for instance scale-free spectra $P(k) \propto
k^{n}$, which yield $\sgw(M) \propto M^{-(n+3)/3}$ where $n\ngt -3$ is
physically plausible. The r.h.s. of equation (\ref{rps}) becomes
negative for $M<M_{\rm c}(t)$ where $M_{\rm c}(t)$ is the critical
mass at which $\sigma(M)=\delc(t)$. This clearly indicates that
$R_{\rm PS}(M,t)$ should not be interpreted as the formation rate of
bound objects.  Rather it corresponds to the net rate of change in the
number density. In other words, $R_{\rm PS}(M,t)$ consists of the
formation rate $R_{\rm form}(M,t)$ at which objects of a given mass
$M$ are formed by mergers of smaller mass objects, and the destruction
rate $R_{\rm dest}(M,t)$ at which they are destroyed by coalescence to
produce more massive systems. Thus equation (\ref{rps}) should
formally be written as
%%%%%%%%%%%%%%%%%%%%%%%%%%%%%%%%%%%%%%%%%%%%%%%%%%%%%%%%%%%
\begin{equation}
   R_{\rm PS}(M,t) \equiv R_{\rm form}(M,t) - R_{\rm dest}(M,t) , 
\label{rate}
\end{equation}
%%%%%%%%%%%%%%%%%%%%%%%%%%%%%%%%%%%%%%%%%%%%%%%%%%%%%%%%%%%
(Cavaliere et al. 1991; Blain \& Longair 1993a,b). The negative
$R_{\rm PS}(M,t)$ at $M<M_{\rm c}(t)$ is then ascribed to an excess of
the destruction rate over the formation rate in this mass range. On
the other hand, the formation rate dominates at $M>M_{\rm c}(t)$
resulting in the positive $R_{\rm PS}(M,t)$. In the case of scale-free
spectra with $n > -3$, the critical mass $M_{\rm c}(t)$ increases
monotonically with time because $M_{\rm c}(t) \propto D(t)^{6/(n+3)}$. 
This is a generic feature in the `bottom-up picture' of hierarchical
structure formation, in which small-scale structures form first and
large-scale structures appear later. Our main task in this paper is to
derive analytical expressions for $R_{\rm form}(M,t)$ and $R_{\rm
dest}(M,t)$. In so doing, we find that a more specific definition of
formation and destruction is essential in separating properly the two
terms in equation (\ref{rate}).

%----------------------------------------------------------------------------
\subsection{Conditional probabilities of merger process}

Merging is one of the most fundamental physical processes for the
formation and destruction of virialized systems. LC developed a method
to treat this process analytically following the idea of Bond et
al. (1991).  Their results are strictly justified only when one adopts
the sharp $k$-space filtering, i.e. the spherical top-hat filtering in
$k$-space, and when $\sigma(M)$ is a monotonically decreasing function
of $M$. They found an analytical expression for the conditional
probability that a point in a universe resides in an object of mass
$M_1\sim M_1+dM_1$ at time $t_1$ provided it becomes part of a larger
object of mass $M_2 (>M_1)$ at later time $t_2 (>t_1)$:
%%%%%%%%%%%%%%%%%%%%%%%%%%%%%%%%%%%%%%%%%%%%%%%%%%%%%%%%%%%%%%%%%%%%%%%%
\begin{eqnarray}
 \lefteqn{ P_1(M_1,t_1|M_2,t_2)dM_1} \nonumber \\
     &=& \hspace{-2mm} \frac{1}{\sqrt{2\pi}}
       \frac{\delta_{\rm c1}-\delta_{\rm c2}}{(\sgw_1-\sgw_2)^{3/2}} 
          \abs{\frac{d\sgw_1}{dM_1}} \exp\lbkt{-\frac{(\delta_{\rm c1} 
          -\delta_{\rm c2})^{2}}{2(\sgw_1-\sgw_2)}}dM_1,
\label{p1}
\end{eqnarray}
%%%%%%%%%%%%%%%%%%%%%%%%%%%%%%%%%%%%%%%%%%%%%%%%%%%%%%%%%%%%%%%%%%%%%%%%
where $\sigma_{i}\equiv\sigma(M_{i})$ and
$\delta_{ci}\equiv\delta_{c}(t_{i})$ for $i=1$ and $2$.  The same
expression was obtained in a more heuristic manner by Bower (1991).
The reverse conditional probability that a point resides in an object
of mass $M_2 \sim M_2+dM_2$ at time $t_2$ provided it has been part of
a smaller object of mass $M_1(<M_2)$ at $t_1 (<t_2)$ is
%%%%%%%%%%%%%%%%%%%%%%%%%%%%%%%%%%%%%%%%%%%%%%%%%%%%%%%%%%%%%%%%%%%%%
\begin{eqnarray}
  \lefteqn{P_2(M_2,t_2|M_1,t_1)dM_2}  \nonumber \\
    &= &  \frac{1}{\sqrt{2\pi}} \frac{\delta_{\rm c2}(\delta_{\rm c1} 
              -\delta_{\rm c2})}{\delta_{\rm c1}}
            \lbkt{\frac{\sgw_1}{\sgw_2(\sgw_1-\sgw_2)}}^{\frac{3}{2}}
           \abs{\frac{d\sgw_2}{dM_2}} \nonumber \\ 
    & &  \times  \exp\lbkt{-\frac{(\sgw_2\delta_{\rm c1} 
         -\sgw_1\delta_{\rm c2})^{2}}{2\sgw_1\sgw_2(\sgw_1-\sgw_2)}}dM_2 .
\label{p2}
\end{eqnarray}
%%%%%%%%%%%%%%%%%%%%%%%%%%%%%%%%%%%%%%%%%%%%%%%%%%%%%%%%%%%%%%%%%%%%%

Equations (\ref{p1}) and (\ref{p2}) readily yield instantaneous
transition rates from a certain mass to another. Setting $t_1=t-\Delta
t$, $t_2=t$ and $M_2=M$ in equation (\ref{p1}), and taking the limit
$\Delta t \rightarrow 0$, one obtains
%%%%%%%%%%%%%%%%%%%%%%%%%%%%%%%%%%%%%%%%%%%%%%%%%%%%%%%%%%%%%%%%%%%%%
\begin{eqnarray}
 \lefteqn{ \frac{dP_1(M_1 \rightarrow M; t)}{dt}dM_1 
    \equiv  \lim_{\Delta t \rightarrow 0}
        \frac{P_1(M_1,t-\Delta t|M,t)}{\Delta t}dM_1 } \nonumber \\
     & = &  \frac{1}{\sqrt{2\pi}} \frac{1}{(\sgw_1-\sigma^{2})^{3/2}}
               \lbkt{-\frac{d\delc(t)}{dt}} \abs{\frac{d\sgw_1}{dM_1}}dM_1 .
\label{t1}
\end{eqnarray}
%%%%%%%%%%%%%%%%%%%%%%%%%%%%%%%%%%%%%%%%%%%%%%%%%%%%%%%%%%%%%%%%%%%%%
This quantity is interpreted as the rate at which an object of mass $M
(>M_1)$ is formed from an object of mass $M_1\sim M_1+dM_1$ in unit
time at $t$.  Similarly from equation (\ref{p2}), one finds
%%%%%%%%%%%%%%%%%%%%%%%%%%%%%%%%%%%%%%%%%%%%%%%%%%%%%%%%%%%%%%%%%%%%%%%%%%%%
\begin{eqnarray}
 \lefteqn{ \frac{dP_2(M \rightarrow M_2; t)}{dt}dM_2 \equiv 
       \lim_{\Delta t \rightarrow 0} 
       \frac{P_2(M_2,t+\Delta t|M,t)}{\Delta t}dM_2} \nonumber \\
      & = &   \frac{1}{\sqrt{2\pi}} 
           \lbkt{\frac{\sigma^{2}}{\sgw_2(\sigma^{2}-\sgw_2)}}^{\frac{3}{2}}
            \lbkt{-\frac{d\delc(t)}{dt}}\abs{\frac{d\sgw_2}{dM_2}}\nonumber \\
      &   &    \times \exp\lbkt{-\frac{(\sigma^{2}-\sgw_2)\delc^{2}(t)}
                                   {2\sigma^{2}\sgw_2}}dM_2 . 
\label{t2}
\end{eqnarray}
%%%%%%%%%%%%%%%%%%%%%%%%%%%%%%%%%%%%%%%%%%%%%%%%%%%%%%%%%%%%%%%%%%%%%%%%%%%%
Equation (\ref{t2}) corresponds to the rate at which an object of mass
$M (<M_2)$ is incorporated into a larger object of mass $M_2 \sim
M_2+dM_2$ at time $t$.

%------------------------------------------------------------------------
\subsection{Formal procedure}

Now we can write down the rates of formation and destruction of
gravitationally bound systems using the transition rates in the last
subsection. From equation (\ref{t1}), one {\it formally} defines the
formation rate of an object of mass $M$ at time $t$ as
%%%%%%%%%%%%%%%%%%%%%%%%%%%%%%%%%%%%%%%%%%%%%%%%%%%%%%%%%%%%%%%%%%%%%
\begin{eqnarray}
  \lefteqn{R_{\rm form}(M,t) \equiv \int_{0}^{M}dM_1
          \frac{dP_1(M_1\rightarrow M; t)}{dt} N_{\rm PS}(M,t)} \nonumber \\
           & = & \lbkt{ \frac{1}{\sqrt{\pi}} \frac{\delc(t)}{\sgw(M)}
                        \int_{0}^{\infty}\frac{dx}{x^{2}}}
                              \lbkt{-\frac{d\delc(t)}{dt}}N_{\rm PS}(M,t) . 
\label{rf}
\end{eqnarray}
%%%%%%%%%%%%%%%%%%%%%%%%%%%%%%%%%%%%%%%%%%%%%%%%%%%%%%%%%%%%%%%%%%%%%
Equation (\ref{t2}) similarly defines the {\it formal} destruction rate:
%%%%%%%%%%%%%%%%%%%%%%%%%%%%%%%%%%%%%%%%%%%%%%%%%%%%%%%%%%%%%%%%%%%%%
\begin{eqnarray}
 \lefteqn{ R_{\rm dest}(M,t) \equiv \int^{\infty}_{M}dM_2
          \frac{dP_2(M \rightarrow M_2; t)}{dt} N_{\rm PS}(M,t)} \nonumber \\
            & = & \lbkt{ \frac{1}{\delc(t)}+\frac{1}{\sqrt{\pi}}
                  \frac{\delc(t)}{\sgw(M)} \int_{0}^{\infty}\frac{e^{-y^{2}}
                    } {y^{2}}dy} \nonumber \\
            &  &     \times \lbkt{-\frac{d\delc(t)}{dt}}N_{\rm PS}(M,t) .
\label{rd}
\end{eqnarray}
%%%%%%%%%%%%%%%%%%%%%%%%%%%%%%%%%%%%%%%%%%%%%%%%%%%%%%%%%%%%%%%%%%%%%
Clearly these formal expressions diverge, reflecting the divergence of
equations (\ref{t1}) and (\ref{t2}) in the limit $M_1\rightarrow M$
and $M_2 \rightarrow M$, respectively.  Such a divergence would simply
be related to the fact that the above formalism identifies the
formation and destruction of objects even if the associate change of
mass is infinitesimal.  In reality, however, such a small change in
mass hardly corresponds to the formation or destruction in a usual
sense; rather it should be regarded as accretion or evolution of the
same system.  In order to incorporate more realistic and physically
reasonable description of these events, we should replace the above
definitions of formation and destruction with more appropriate ones.
Then the divergence associated in equations (\ref{rf}) and (\ref{rd})
would disappear naturally.

One may argue that the divergence is ascribed to counting the
transition of a certain object to practically the {\it same} one, by
showing explicitly that equations (\ref{rf}) and (\ref{rd}) diverge in
a similar manner. In fact, one can prove that the divergences in the
two rates exactly cancel out and their difference reduces to the time
derivative of the PS mass function:
%%%%%%%%%%%%%%%%%%%%%%%%%%%%%%%%%%%%%%%%%%%%%%%%%%%%%%%%%%%%%%%%%%%%%
\begin{eqnarray}
  \lefteqn{R_{\rm form}(M,t) -R_{\rm dest}(M,t)} \nonumber \\ 
         & = & \lbkt{\frac{\delc(t)}{\sgw(M)}-\frac{1}{\delc(t)}}
                \lbkt{-\frac{d\delc(t)}{dt}}N_{\rm PS}(M,t) \nonumber \\
         & = & R_{\rm PS}(M,t) .  
\label{comb}
\end{eqnarray}
%%%%%%%%%%%%%%%%%%%%%%%%%%%%%%%%%%%%%%%%%%%%%%%%%%%%%%%%%%%%%%%%%%%%%
The above result also suggests that the divergence in equations
(\ref{rf}) and (\ref{rd}) can be removed by redefining the formation
and destruction events properly. We will propose a realistic and
reasonable prescription in the next subsection.

\hfill
%---------------------------------------------------------------------------
\subsection{Realistic prescription to compute 
$R_{\rm form}$ and $R_{\rm dest}$}

Since the divergence mentioned in the last subsection is originated
from counting the transitions from an object to the essentially {\it
same} one, a simplest remedy is to introduce thresholds in the mass of
formation and destruction events.  More specifically, we replace
equations (\ref{rf}) and (\ref{rd}) with
%%%%%%%%%%%%%%%%%%%%%%%%%%%%%%%%%%%%%%%%%%%%%%%%%%%%%%%%%%%%%%%%%%%%%
\begin{equation}
  R_{\rm form}(M,t;M_{\rm f}) \equiv \int_{0}^{M_{\rm f}} \hspace{-2mm}dM_1
                       \frac{dP_1(M_1\rightarrow M;t)}{dt} N_{\rm PS}(M,t) , 
\label{formdef} 
\end{equation}
%%%%%%%%%%%%%%%%%%%%%%%%%%%%%%%%%%%%%%%%%%%%%%%%%%%%%%%%%%%%%%%%%%%%%
and
%%%%%%%%%%%%%%%%%%%%%%%%%%%%%%%%%%%%%%%%%%%%%%%%%%%%%%%%%%%%%%%%%%%%%
\begin{equation}
  R_{\rm dest}(M,t;M_{\rm d}) \equiv \int^{\infty}_{M_{\rm d}}\hspace{-1mm}dM_2
                      \frac{dP_2(M \rightarrow M_2;t)}{dt} N_{\rm PS}(M,t) .
\label{destdef}
\end{equation}
%%%%%%%%%%%%%%%%%%%%%%%%%%%%%%%%%%%%%%%%%%%%%%%%%%%%%%%%%%%%%%%%%%%%%
Physically the above procedure corresponds to assuming that an object
of mass $M$ keeps its identity as long as it stays between $M_{\rm f}$ and
$M_{\rm d}$, and to defining the formation and destruction of the object
only when the associate mass change is out of this range. Adopting
such a prescription, the above formulae are integrated to give
%%%%%%%%%%%%%%%%%%%%%%%%%%%%%%%%%%%%%%%%%%%%%%%%%%%%%%%%%%%%%%%%%%%%%
\begin{eqnarray}
  \lefteqn{R_{\rm form}(M,t;M_{\rm f})}  \nonumber \\
     &=&   \sqrt{\frac{2}{\pi}} 
                            \frac{1}{\sqrt{\sgw(M_{\rm f})-\sgw(M)}}
                            \lbkt{-\frac{d\delc(t)}{dt}}N_{\rm PS}(M,t),
\label{form}
\end{eqnarray}
%%%%%%%%%%%%%%%%%%%%%%%%%%%%%%%%%%%%%%%%%%%%%%%%%%%%%%%%%%%%%%%%%%%%%
and
%%%%%%%%%%%%%%%%%%%%%%%%%%%%%%%%%%%%%%%%%%%%%%%%%%%%%%%%%%%%%%%%%%%%%
\begin{eqnarray}
  \lefteqn{R_{\rm dest}(M,t;M_{\rm d})} \nonumber \\
  &=& \lbkt{\sbkt{\frac{1}{\delc(t)}-\frac{\delc(t)}{\sgw(M)}}
                 {\rm erfc}(Z)+\frac{1}{\sqrt{\pi}}\frac{\delc(t)}{\sgw(M)}
                  \frac{e^{-Z^{2}}}{Z}} \nonumber \\
  & & \times  \lbkt{-\frac{d\delc(t)}{dt}}N_{\rm PS}(M,t),
\label{dest}
\end{eqnarray}
%%%%%%%%%%%%%%%%%%%%%%%%%%%%%%%%%%%%%%%%%%%%%%%%%%%%%%%%%%%%%%%%%%%%%
where $Z$ is defined as 
%%%%%%%%%%%%%%%%%%%%%%%%%%%%%%%%%%%%%%%%%%%%%%%%%%%%%%%%%%%%%%%%%%%%%
\begin{equation}        
  Z(M,t;M_{\rm d}) \equiv \delc(t)\sqrt{\frac{\sgw(M) 
             - \sgw(M_{\rm d})}{2\sgw(M)\sgw(M_{\rm d})}},
\end{equation}
%%%%%%%%%%%%%%%%%%%%%%%%%%%%%%%%%%%%%%%%%%%%%%%%%%%%%%%%%%%%%%%%%%%%%
and erfc($u$) is the complimentary error function:
%%%%%%%%%%%%%%%%%%%%%%%%%%%%%%%%%%%%%%%%%%%%%%%%%%%%%%%%%%%%%%%%%%%%%
\begin{equation}
    {\rm erfc}(u)\equiv 
       \frac{2}{\sqrt{\pi}}\int_{u}^{\infty}\exp(-t^{2})dt.
\end{equation}
%%%%%%%%%%%%%%%%%%%%%%%%%%%%%%%%%%%%%%%%%%%%%%%%%%%%%%%%%%%%%%%%%%%%%

In the above prescription, it is important to choose physically
reasonable values for $M_{\rm f}$ and $M_{\rm d}$.  Admittedly it is
inevitably artificial. We here propose to set $M_{\rm f}=M/2$ and
$M_{\rm d}=2M$ as in LC mainly because of the
following two reasons. One is that the identity of an object of mass
$M$ is clearly retained between $M/2$ and $2M$.  If a new object of
mass $M$ were formed from a parent object of mass greater than $M/2$,
the parent object would make up the dominant part of the new
object. Therefore such a process should correspond to {\it evolution}
of the parent object rather than formation of a new object. The same
is true for the destruction of an object of mass $M$ which is
incorporated into an object of mass less than $2M$. 

%##################################################################
\begin{figure}
\begin{center}
   \leavevmode\psfig{figure=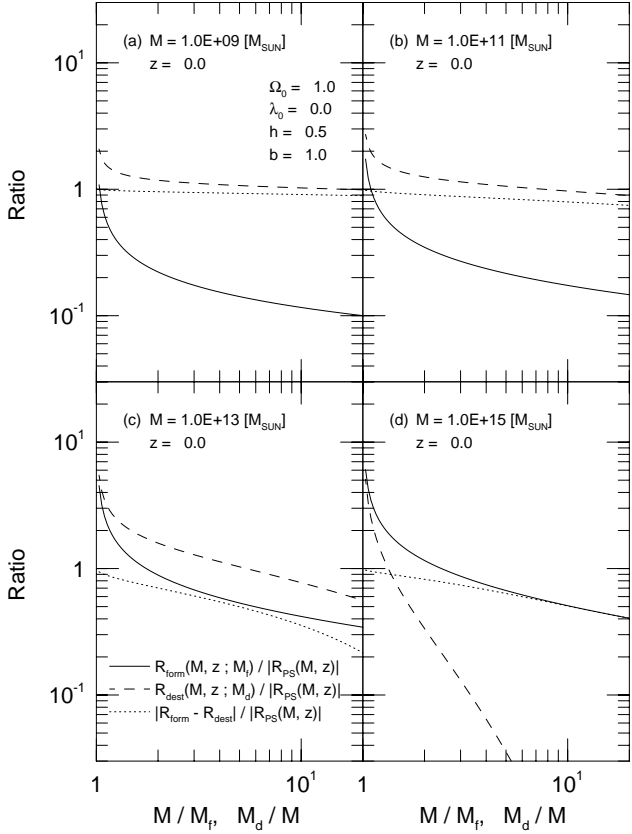,height=11cm}
\end{center}
\label{figratio}
\caption{Dependence of $R_{\rm form}(M,z;M_{\rm f})$ (solid curve),
$R_{\rm dest}(M,z;M_{\rm d})$ (dashed), and their difference (dotted)
on $M/M_{\rm f}$ or $M_{\rm d}/M$ at $z=0$ in the standard CDM model
(each quantity is normalized by $|R_{\rm PS}(M,z)|$); (a) $M=10^9
\msun$, (b) $M=10^{11} \msun$, (c) $M=10^{13} \msun$, (d) $M=10^{15}
\msun$.  }
\end{figure}
%##################################################################
The other reason is more practical; equations (\ref{form}) and
(\ref{dest}) are rather insensitive to the specific choice of $M_{\rm
f}$ and $M_{\rm d}$ at $M_{\rm f}\approx M/2$ and $M_{\rm d}\approx
2M$, and their difference agrees fairly well with $R_{\rm
PS}(M,t)$. To see such features more clearly, we plot $R_{\rm
form}(M,z;M_{\rm f})$, $R_{\rm dest}(M,z;M_{\rm d})$ and their
difference against $M/M_{\rm f}$ or $M_{\rm d}/M$ in Fig.~1 (each
quantity is normalized by $|R_{\rm PS}(M,z)|$). Hereafter in this
paper, a CDM power spectrum of Davis et al. (1985) is adopted and the
mass variance is evaluated by the following approximation (White \&
Frenk 1991):
%%%%%%%%%%%%%%%%%%%%%%%%%%%%%%%%%%%%%%%%%%%%%%%%%%%%%%%%%%%%%%%%%%%%%
\begin{eqnarray}
\sigma &\propto& \left[ 1-0.4490 (\Omega_0 h)^{0.1}
        \sbkt{\frac{r_{\rm f}}{h^{-1}\rm Mpc}}^{0.1} \right. \nonumber \\
   & & + \left. 0.6352 (\Omega_0 h)^{0.2}\sbkt{\frac{r_{\rm f}}{h^{-1}\rm
            Mpc}}^{0.2}\right]^{-10} , 
\label{cdm}
\end{eqnarray}
%%%%%%%%%%%%%%%%%%%%%%%%%%%%%%%%%%%%%%%%%%%%%%%%%%%%%%%%%%%%%%%%%%%%%
where $h\equiv H_0/(\mbox{100 km/sec/Mpc})$ is the dimensionless
Hubble constant, and the top-hat filtering is implicitly assumed. The
above approximation holds within 10\% over the range $0.013(\Omega_0
h)^{-1} < r_{\rm f}/(h^{-1}{\rm Mpc}) < 10 (\Omega_0
h)^{-1}$. Equation (\ref{cdm}) is normalized by the relation
$\sigma(r_{\rm f}=8h^{-1}{\rm Mpc})=b^{-1}$, where $b$ is the biasing
parameter.  In this section, we present results only in the standard
CDM model ($\Omega_0=1$, $h=0.5$, and $b=1$) for an illustrative
purpose, while we consider other CDM variants (e.g., Table 1) in
Section 3.

Panels (a)-(d) in Fig.~1 are plotted for several masses at present
redshift $z=0$. The figure shows that $R_{\rm form}/|R_{\rm PS}|$ and
$R_{\rm dest}/|R_{\rm PS}|$ are fairly insensitive to $M/M_{\rm f}$ or
$M_{\rm d}/M$ for $M/M_{\rm f},M_{\rm d}/M$ \ngt $1.5$, and that
$R_{\rm form}-R_{\rm dest} \approx R_{\rm PS}$ within $30\%$ for
$M/M_{\rm f},M_{\rm d}/M$ \nlt $3$. Thus for practical purposes as
well, the choice of $M_{\rm f}=M/2$ and $M_{\rm d}=2M$ seems
reasonable.

%##################################################################
\begin{figure*}
\begin{center}
   \leavevmode\psfig{figure=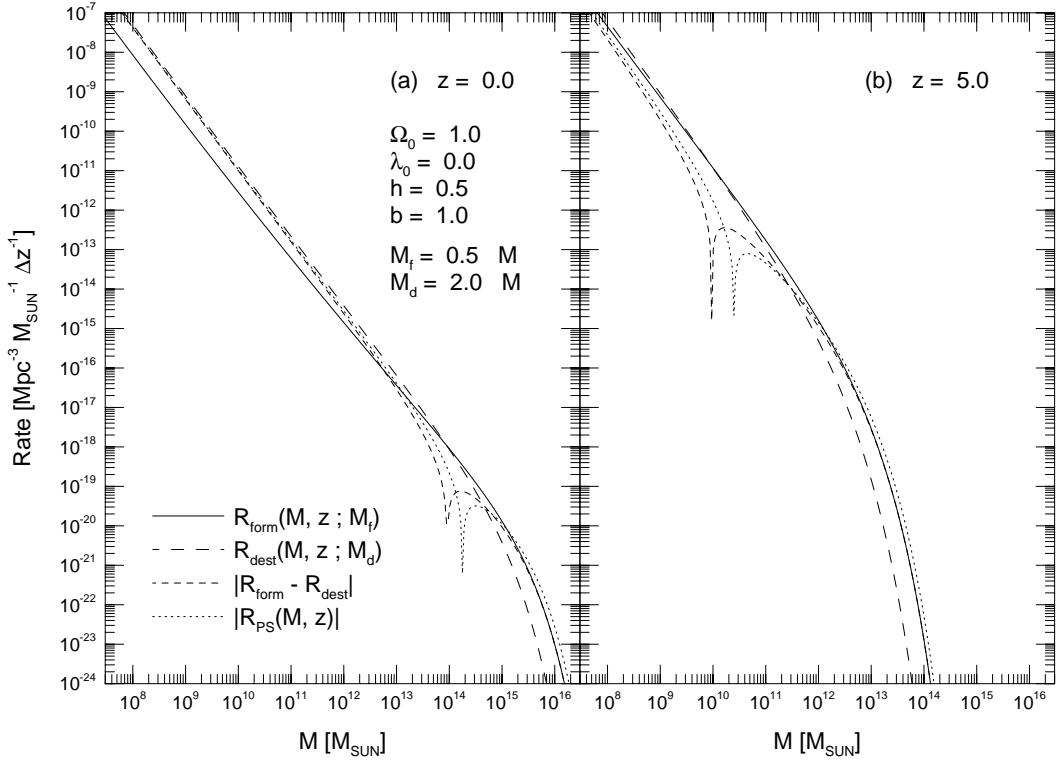,height=10cm,angle=90}
\end{center}
\caption{Rates of formation $R_{\rm form}(M,z,M_{\rm f})$ (solid
curve) and destruction $R_{\rm dest}(M,z,M_{\rm d})$ (long-dashed) in
the case of $M_{\rm f}=M/2$ and $M_{\rm d}=2M$, as a function of mass
in the standard CDM model. Also plotted are the absolute values of
their difference (short-dashed) and $R_{\rm PS}(M,z)$ (dotted); (a)
$z=0$, (b) $z=5$.  }
\end{figure*}
%##################################################################
Figure~2 illustrates $R_{\rm form}(M,z;M/2)$, $R_{\rm dest}(M,z;2M)$,
their difference, and $|R_{\rm PS}(M,z)|$ versus $M$ in the standard
CDM model. Both panels exhibit that $R_{\rm form}<R_{\rm dest}$ for
small masses and $R_{\rm form}>R_{\rm dest}$ for large masses. The
critical mass $M_{\rm c}$ at which $R_{\rm form}=R_{\rm dest}$
increases with time; $M_{\rm c}\approx 10^{10}\msun$ at $z=5$, and
$M_{\rm c}\approx 10^{14}\msun$ at $z=0$ in this example.  The
difference between two rates is very close to $R_{\rm PS}$ except at
$M \approx M_{\rm c}$. The deviation simply arises from the fact that
we set $M_{\rm f}, M_{\rm d} \not= M$, but is not important in
practice because its amplitude is several orders of magnitude smaller
than those of $R_{\rm form}$ and $R_{\rm dest}$.  With these results
in mind, we adopt $M_{\rm f}=M/2$ and $M_{\rm d}=2M$ in the rest of
this paper, and simply use $R_{\rm form}(M,t)$ and $R_{\rm dest}(M,t)$
to denote $R_{\rm form}(M,t;M/2)$ and $R_{\rm dest}(M,t;2M)$
respectively.

%\hfill
%---------------------------------------------------------------------------
\subsection{Formation epoch distribution} 

Before proceeding to the application of the results obtained in the
previous subsections, let us discuss the formation epoch of virialized
systems in some detail. First consider the probability that an object
of mass $M$ at time $t_1$ remains, without destructed, to have mass
less than $2M$ at later time $t_2(>t_1)$.  We call this quantity {\it
the survival probability}. It is evaluated by LC from equation
(\ref{p2}) as 
%%%%%%%%%%%%%%%%%%%%%%%%%%%%%%%%%%%%%%%%%%%%%%%%%%%%%%%%%%%%%%%%%%%%%
\begin{eqnarray}
 \lefteqn{P_{\rm surv}(M,t_1,t_2) \equiv \int^{2M}_{M} P_2(M_2,t_2|
    M,t_1)dM_2} \nonumber \\ &=&\hspace{-2mm} 1-
    \frac{1}{2}\lbkt{\frac{\delta_{\rm c1}-2\delta_{\rm c2}}
    {\delta_{\rm c1}}}\exp\lbkt{\frac{2\delta_{\rm c2} (\delta_{\rm
    c1}-\delta_{\rm c2})}{\sgw(M)}} \nonumber \\ & & \times {\rm
    erfc}[X(M,t_1,t_2)] -\frac{1}{2}{\rm erfc}[Y(M,t_1,t_2)],
\label{surv}
\end{eqnarray}
%%%%%%%%%%%%%%%%%%%%%%%%%%%%%%%%%%%%%%%%%%%%%%%%%%%%%%%%%%%%%%%%%%%%%
where $X(M,t_1,t_2)$ and $Y(M,t_1,t_2)$ are defined respectively as 
%%%%%%%%%%%%%%%%%%%%%%%%%%%%%%%%%%%%%%%%%%%%%%%%%%%%%%%%%%%%%%%%%%%%%
\begin{equation}
   X(M,t_1,t_2)\equiv\frac{\sigma^{2}(2M)
       [\delta_{\rm c1}-2\delta_{\rm c2}]+\sgw(M)\delta_{\rm c2}}
               {\sqrt{2\sgw(M)\sigma^{2}(2M)[\sgw(M)-\sigma^{2}(2M)]}},
\end{equation}
%%%%%%%%%%%%%%%%%%%%%%%%%%%%%%%%%%%%%%%%%%%%%%%%%%%%%%%%%%%%%%%%%%%%%
and
%%%%%%%%%%%%%%%%%%%%%%%%%%%%%%%%%%%%%%%%%%%%%%%%%%%%%%%%%%%%%%%%%%%%%
\begin{equation}
   Y(M,t_1,t_2)\equiv
        \frac{\sgw(M)\delta_{\rm c2}-\sigma^{2}(2M)\delta_{\rm c1}}
                 {\sqrt{2\sgw(M)\sigma^{2}(2M)[\sgw(M)-\sigma^{2}(2M)]}}.
\end{equation}
%%%%%%%%%%%%%%%%%%%%%%%%%%%%%%%%%%%%%%%%%%%%%%%%%%%%%%%%%%%%%%%%%%%%%

Using the above formulae and equation (\ref{form}), we write down the
number density of bound systems which form with mass $M\sim M+dM$ at
time $t_{\rm f} \sim t_{\rm f}+dt_{\rm f}$ and survive without
destructed until later time $t$ as
%%%%%%%%%%%%%%%%%%%%%%%%%%%%%%%%%%%%%%%%%%%%%%%%%%%%%%%%%%%%%%%%%%%%%
\begin{eqnarray}
 \lefteqn{F(M,t_{\rm f},t)dMdt_{\rm f} \equiv 
          R_{\rm form}(M,t_{\rm f}) dM dt_{\rm f} \times 
          P_{\rm surv}(M,t_{\rm f},t)}\nonumber \\ 
    &=&\frac{1}{\sqrt{2\pi}}\frac{1}{\sqrt{\sigma^{2}(M/2)-\sgw(M)}} 
        \left[2-\sbkt{\frac{\delc(t_{\rm f})-2\delc(t)}{\delc(t)}}
             \right.\nonumber \\ 
    & & \times \exp\sbkt{\frac{2\delc(t)[\delc(t_{\rm f})-\delc(t)]} 
          {\sgw(M)}} {\rm erfc}[X(M,t_{\rm f},t)] \nonumber \\ 
     & & - {\rm erfc}[Y(M,t_{\rm f},t)]
             \left.\begin{array}{c}\\\\\end{array}\hspace{-4mm}\right]
             \lbkt{-\frac{d\delc}{dt} (t_{\rm f})}
              N_{\rm PS}(M,t_{\rm f})dM dt_{\rm f} .
\label{fdist}
\end{eqnarray}
%%%%%%%%%%%%%%%%%%%%%%%%%%%%%%%%%%%%%%%%%%%%%%%%%%%%%%%%%%%%%%%%%%%%%
Given a mass $M$ and time $t$, equation (\ref{fdist}) is regarded as
the distribution function of formation epoch $t_{\rm f}$.
%##################################################################
\begin{figure}
\begin{center}
   \leavevmode\psfig{figure=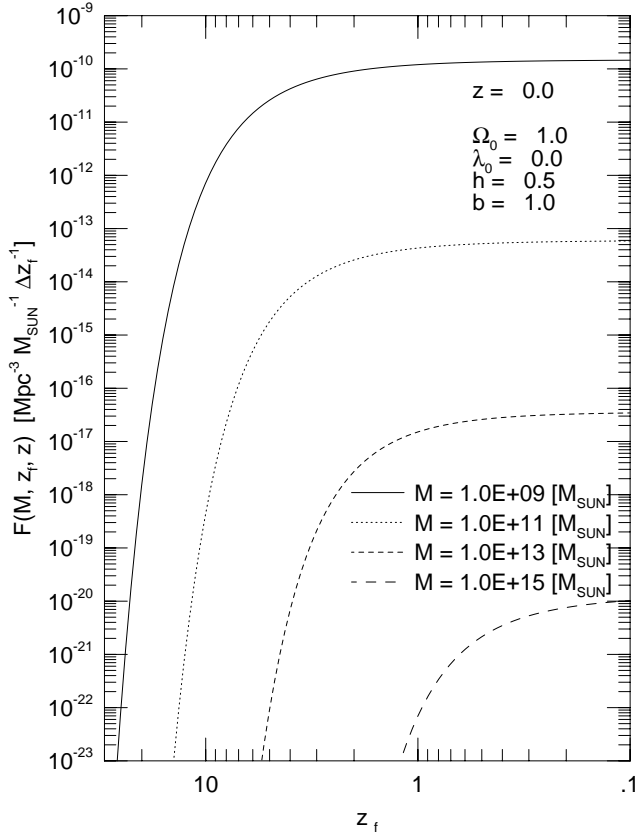,height=11cm}
\end{center}
\caption{Distribution of formation epoch $F(M,z_{\rm f},z)$ in the
standard CDM model. The differential number densities of bound objects
that are formed at redshift $z_{\rm f}$ and survive until $z=0$ are
plotted against $z_{\rm f}$ for masses $M=10^9 \msun$ (solid curve),
$10^{11} \msun$ (dotted), $10^{13} \msun$ (short-dashed) and $10^{15}
\msun$ (long-dashed). }
\end{figure}
%##################################################################
In Fig.~3 we plot $F(M,z_{\rm f},z=0)$ versus $z_{\rm f}$ for several
masses in the standard CDM model. A common feature is that $F$ as a
function of $z_{\rm f}$ remains nearly constant up to a certain
redshift and then starts to decline. The epoch at which $F$ drops by
an order of magnitude is $z_{\rm f}\approx 6$ for $M=10^{9}\msun$,
$z_{\rm f}\approx 4$ for $M=10^{11}\msun$, $z_{\rm f}\approx 2$ for
$M=10^{13}\msun$, and $z_{\rm f}\approx 0.6$ for
$M=10^{15}\msun$. These are interpreted as typical epochs for the
formation of bound systems of respective masses which exist at $z=0$.

\section{X-RAY BACKGROUND RADIATION CONTRIBUTED FROM CLUSTERS OF GALAXIES}

As a simple and important application of our formulae, we calculate
X-ray emission from clusters of galaxies and their contribution to the
observed XRB. The formation rate of virialized systems plays an
essential role in specifying the epoch of formation of clusters and in
explicitly taking account of their luminosity evolution.

%%%%%%%%%%%%%%%%%%%%%%%%%%%%%%%%%%%%%%%%%%%%%%%%%%%%%%%%%%%%%%%%%%%%%
\subsection{Method of calculation}

\subsubsection{Cluster contribution to the XRB spectra}
\label{xrb}
Clusters of galaxies are filled with diffuse hot plasma which produces
copious X-ray emission. If the total power liberated per unit comoving
volume and unit frequency at redshift $z$ is denoted by $J_{\nu}(z)$,
the background X-ray intensity $I_{\nu_0}$ observed at present at
frequency $\nu_0$ is described as
%%%%%%%%%%%%%%%%%%%%%%%%%%%%%%%%%%%%%%%%%%%%%%%%%%%%%%%%%%%%%%%%%%%%%%%%
\begin{equation}
I_{\nu_0}=\frac{c}{4\pi H_0}\int_0^{\infty} \hspace{-3mm}
\frac{J_{\nu_0(1+z)}(z) \; dz} {(1+z)\sqrt{(\Omega_0 z+1)(1+z)^2 -
\lambda_0 z(2+z)}} ,
\label{spectrum}
\end{equation}
%%%%%%%%%%%%%%%%%%%%%%%%%%%%%%%%%%%%%%%%%%%%%%%%%%%%%%%%%%%%%%%%%%%%%%%%
where $c$ is the speed of light, and $\lambda_0$ is the dimensionless
cosmological constant.  In our model, the luminosity at frequency
$\nu$ of a cluster, $L_{\nu}(M,z_{\rm f},z)$, is assumed to be
determined by the mass $M$, the epoch of its formation $z_{\rm f}$,
and the epoch $z$ at which the observed X-ray is emitted (see
eq.[\ref{luminosity}] below). Then $J_{\nu}(z)$ is given by
%%%%%%%%%%%%%%%%%%%%%%%%%%%%%%%%%%%%%%%%%%%%%%%%%%%%%%%%%%%%%%%%%%%%%%%%
\begin{equation}
J_{\nu}(z)=  \int_0^{\infty}dM \int_{z}^{\infty}dz_{\rm f}
  F(M,z_{\rm f},z) L_{\nu}(M,z_{\rm f},z),
\label{power}
\end{equation}
%%%%%%%%%%%%%%%%%%%%%%%%%%%%%%%%%%%%%%%%%%%%%%%%%%%%%%%%%%%%%%%%%%%%%%%%
where $F(M,z_{\rm f},z)$ is the formation epoch distribution
(eq.~[\ref{fdist}]).

In contrast to equation (\ref{power}), previous authors calculated
$J_{\nu}(z)$ using the mass function of clusters directly as follows
(Evrard \& Henry 1991; Blanchard et al. 1992; Burg et al 1993):
%%%%%%%%%%%%%%%%%%%%%%%%%%%%%%%%%%%%%%%%%%%%%%%%%%%%%%%%%%%%%%%%%%%%%%%%
\begin{equation}
J_{\nu}(z) = \int_0^{\infty} dM N(M,z) L_{\nu}(M,z_{\rm f}=z,z), 
\label{power2}
\end{equation}
%%%%%%%%%%%%%%%%%%%%%%%%%%%%%%%%%%%%%%%%%%%%%%%%%%%%%%%%%%%%%%%%%%%%%%%% 
where the mass function $N(M,z)$ is often taken as the PS formula
(eq.~[\ref{ps}]). Since the mass function does not specify the epoch
of formation of virialized clusters, $z_{\rm f}$ is usually set to be
equal to the epoch of emission $z$. Such a treatment, however, is
strictly justified only if $z_{\rm f}$ is very close to $z$, and will
become inappropriate otherwise. The properties of a virialized system
such as radius and temperature are determined by $z_{\rm f}$ not $z$
(see Section \ref{xray}). Thus the luminosity evolution described in
equation (\ref{power2}) is not realistic. In fact one frequently
encounters similar situations where the formation epoch is essential
in incorporating the evolution of individual astronomical objects and
making cosmological predictions. This is one of the strong advantages
of our current refinement of the original PS theory.

\subsubsection{X-ray emission from clusters}
\label{xray}

The total gravitating mass of clusters is dominated by dissipationless
dark matter, which is likely to be relaxed immediately after the
collapse of halo and attain the virial equilibrium. In a theoretical
treatment for the collapse of spherical perturbations, the radius of a
virialized cluster in the Einstein-de Sitter ($\Omega_0=1$) universe
is estimated as a half of maximum turn-around radius and given by
%%%%%%%%%%%%%%%%%%%%%%%%%%%%%%%%%%%%%%%%%%%%%%%%%%%%%%%%%%%%%%%%%%%%%%%%
\begin{eqnarray} 
r_{\rm vir}&=&\frac{(GM)^{1/3}}{(3\pi H_0)^{2/3}}
                    \frac{1}{1+z_{\rm f}} \nonumber \\
      &=& 1.69(1+z_{\rm f})^{-1} \sbkt{\frac{M}{10^{15}\msun}}^{1/3} 
              h^{-2/3} \mbox{  Mpc}, 
\label{virrad}
\end{eqnarray}
%%%%%%%%%%%%%%%%%%%%%%%%%%%%%%%%%%%%%%%%%%%%%%%%%%%%%%%%%%%%%%%%%%%%%%%%
where $M$ is the total mass of the cluster, and $z_{\rm f}$ is the
redshift of its formation. The virial temperature of the system
$T_{\rm vir}$ is then calculated as
%%%%%%%%%%%%%%%%%%%%%%%%%%%%%%%%%%%%%%%%%%%%%%%%%%%%%%%%%%%%%%%%%%%%%%%
\begin{eqnarray} 
\lefteqn{T_{\rm vir} = \frac{G M \mu m_{\rm p}}{3k_{\rm B} 
                    r_{\rm vir}}} \nonumber \\
&=&\hspace{-2mm}6.06 \times 10^{7} \sbkt{\frac{\mu}{0.59}}
(1+z_{\rm f}) \sbkt{\frac{M}{10^{15}\msun}}^{2/3} h^{2/3} \mbox{  K}, 
\label{virtemp}
\end{eqnarray} 
%%%%%%%%%%%%%%%%%%%%%%%%%%%%%%%%%%%%%%%%%%%%%%%%%%%%%%%%%%%%%%%%%%%%%%%%
where $G$ is the gravitational constant, $k_{\rm B}$ is the Boltzmann
constant, $m_{\rm p}$ is the proton mass, and $\mu$ is the mean
molecular weight. The cases of $\Omega_0 < 1$ universes are discussed
in Appendix A. Hereafter we assume that the intracluster gas is fully
ionized with the primordial abundance $Y=0.24$, where $Y$ is the
helium fraction by weight, and thus $\mu = 0.59$. Note that once a
model for the background universe is specified, properties of
virialized clusters are determined by two parameters; mass $M$ and the
formation epoch $z_{\rm f}$ which is not available in the conventional
PS theory.

The gas component of clusters generates the X-ray continuum 
emission mainly due to thermal bremsstrahlung
whose emissivity in CGS units is (e.g., Rybicki \& Lightman 1979)
%%%%%%%%%%%%%%%%%%%%%%%%%%%%%%%%%%%%%%%%%%%%%%%%%%%%%%%%%%%%%%%%%%%%%%%%
\begin{eqnarray}
\epsilon_{\nu}^{\rm f\,f}
&=&6.84\times 10^{-38} \sbkt{\frac{2}{2-Y}} \bar{g}(T_{\rm gas},\nu)
\nonumber \\
& & \hspace{-2mm} \times n_{\rm e}^2 T_{\rm gas}^{-1/2}
\exp\sbkt{-\frac{h\nu}{k_{\rm B}T_{\rm gas}}}
\mbox{ \hspace{2mm} erg\,s$^{-1}$\,cm$^{-3}$\,Hz$^{-1}$},
\label{bremss}
\end{eqnarray}
%%%%%%%%%%%%%%%%%%%%%%%%%%%%%%%%%%%%%%%%%%%%%%%%%%%%%%%%%%%%%%%%%%%%%%%%
where $h$ is the Planck constant, $T_{\rm gas}$ is the gas
temperature, and $n_{\rm e}$ is the electron density.  We adopted an
approximation $\bar{g}(T, \nu)=0.9(h\nu/k_{\rm B}T)^{-0.3}$ for the
velocity-averaged Gaunt factor following Evrard \& Henry (1991).

The X-ray emission described by equation (\ref{bremss}) is essentially
determined by the temperature and density profiles of intracluster
gas. Simple models for the intracluster gas often assume that the gas
is isothermal and its distribution is spherically symmetric.  The
observed X-ray spectra and galaxy velocity dispersions further suggest
that $T_{\rm vir}/T_{\rm gas}\approx 1$ (Mushotzky 1984). Thus we set
%%%%%%%%%%%%%%%%%%%%%%%%%%%%%%%%%%%%%%%%%%%%%%%%%%%%%%%%%%%%%%%%%%%%%%%%
\begin{equation}
T_{\rm gas} = T_{\rm vir}(M,z_{\rm f}).
\label{tempprof}
\end{equation}
%%%%%%%%%%%%%%%%%%%%%%%%%%%%%%%%%%%%%%%%%%%%%%%%%%%%%%%%%%%%%%%%%%%%%%%%
For the density profile, the observed X-ray image of surface
brightness is found to agree well with (Jones \& Forman 1984)
%%%%%%%%%%%%%%%%%%%%%%%%%%%%%%%%%%%%%%%%%%%%%%%%%%%%%%%%%%%%%%%%%%%%%%%%
\begin{equation}
\rho_{\rm gas}(r)=\rho_{\rm gas}^0 
                      \lbkt{1+\sbkt{\frac{r}{r_{\rm c}}}^2}^{-1},  
\label{densprof}
\end{equation}
%%%%%%%%%%%%%%%%%%%%%%%%%%%%%%%%%%%%%%%%%%%%%%%%%%%%%%%%%%%%%%%%%%%%%%%%
where $\rho_{\rm gas}^0$ is the central gas density, $r$ is the
distance from the cluster centre, and $r_{\rm c}$ is the core
radius. The above modeling is slightly different from the conventional
isothermal $\beta$-model, in which the temperature and density
profiles of intracluster gas are described by a single parameter
$\beta$.  Although our parametrization seems inconsistent with the
$\beta$ model, it can be reconciled if the galaxy number density
scales as $[1+(r/r_{\rm c})^2]^{-1}$ unlike the King model
$[1+(r/r_{\rm c})^2]^{-3/2}$. In fact, this is qualitatively indicated
from some observations (Bahcall \& Lubin 1994).

As for the core radius $r_{\rm c}$, we still lack understandings of
its physical meaning, and are unable to predict how it depends on $M$
and $z_{\rm f}$.  Thus we adopt a simple self-similar model in which
$r_{\rm c}$ is proportional to $r_{\rm vir}$:
%%%%%%%%%%%%%%%%%%%%%%%%%%%%%%%%%%%%%%%%%%%%%%%%%%%%%%%%%%%%%%%%%%%%%%%%
\begin{eqnarray} 
r_{\rm c} &=& 0.15 h^{-1}\mbox{Mpc } \times \frac{r_{\rm vir}(M,z_{\rm
f})}{r_{\rm vir}(10^{15}\msun, z_{\rm f}=0)} \nonumber \\ 
&=& 0.15(1+z_{\rm f})^{-1} \sbkt{\frac{M}{10^{15}\msun}}^{1/3} 
h^{-1} \mbox{Mpc},
\end{eqnarray}
%%%%%%%%%%%%%%%%%%%%%%%%%%%%%%%%%%%%%%%%%%%%%%%%%%%%%%%%%%%%%%%%%%%%%%%%
where the normalization is chosen to match the {\it Ginga}
observations (Hatsukade 1989), and the second equality assumes the
Einstein-de Sitter universe.  Then the central gas density $\rho_{\rm
gas}^0$ is fixed by
%%%%%%%%%%%%%%%%%%%%%%%%%%%%%%%%%%%%%%%%%%%%%%%%%%%%%%%%%%%%%%%%%%%%%%%%
\begin{equation}
\int_0^{r_{\rm vir}} \rho_{\rm gas}(r)\, 4\pi r^2 dr 
= M \sbkt{\frac{\Omega_{\rm B}}{\Omega_0}} , 
\end{equation}
%%%%%%%%%%%%%%%%%%%%%%%%%%%%%%%%%%%%%%%%%%%%%%%%%%%%%%%%%%%%%%%%%%%%%%%%
where $\Omega_{\rm B}$ is the baryonic density parameter. 

In order to describe realistic X-ray emission from clusters, we also
need to take account of intrinsic luminosity evolution. As the effects
of merger are already factored out in equation (\ref{power}), quiet
gas accretion and the subsequent changes in temperature and density
profiles are here to be incorporated. These processes, however, are
still highly uncertain. So we attempt to model their effects into a
factor $\gamma(z_{\rm f},z)$ and express the luminosity as
%%%%%%%%%%%%%%%%%%%%%%%%%%%%%%%%%%%%%%%%%%%%%%%%%%%%%%%%%%%%%%%%%%%%%% 
\begin{equation} 
L_{\nu}(M,z_{\rm f},z)=\gamma(z_{\rm f},z)\int_0^{r_{\rm vir}}
\epsilon_{\nu}^{\rm f\,f}(M, z_{\rm f}; r)\, 4 \pi r^2 dr ,
\label{luminosity}
\end{equation} 
%%%%%%%%%%%%%%%%%%%%%%%%%%%%%%%%%%%%%%%%%%%%%%%%%%%%%%%%%%%%%%%%%%%%%%%%
where $L_{\nu}(M,z_{\rm f},z)$ is the luminosity that a cluster formed
with mass $M$ at redshift $z_{\rm f}$ would have at a later epoch $z$. 
Given a large degree of uncertainty, $\gamma(z_{\rm f},z)$ is assumed here
to take a power law form:
%%%%%%%%%%%%%%%%%%%%%%%%%%%%%%%%%%%%%%%%%%%%%%%%%%%%%%%%%%%%%%%%%%%%%%%%
\begin{equation} 
\gamma(z_{\rm f},z)=\sbkt{\frac{1+z_{\rm f}}{1+z}}^p.
\label{gamma}
\end{equation}
%%%%%%%%%%%%%%%%%%%%%%%%%%%%%%%%%%%%%%%%%%%%%%%%%%%%%%%%%%%%%%%%%%%%%%%%
Current numerical simulations seem to indicate $p \nlt 1$ (e.g.,
Evrard 1990; Suginohara 1995; Navarro, Frenk \& White 1995), and we
examine two limiting cases of no intrinsic evolution, $p=0$, and of
very strong evolution, $p=1$.

Integrating equation (\ref{luminosity}) over whole frequency, one finds a
simple scaling law for the bolometric luminosity of the form:
%%%%%%%%%%%%%%%%%%%%%%%%%%%%%%%%%%%%%%%%%%%%%%%%%%%%%%%%%%%%%%%%%%%%%%%%
\begin{equation} 
L_{\rm bol}(M,z_{\rm f},z) \propto M^{4/3} (1+z_{\rm f})^{q+p} (1+z)^{-p},
\label{bollum}
\end{equation} 
%%%%%%%%%%%%%%%%%%%%%%%%%%%%%%%%%%%%%%%%%%%%%%%%%%%%%%%%%%%%%%%%%%%%%%%%
where the parameter $q$ is constant and equal to $7/2$ in the
Einstein-de Sitter universe, while it varies with $z_{\rm f}$ and less than
$7/2$ in low $\Omega_0$ universes; e.g. $1.8 \nlt q \nlt 3.5$ for 
$\Omega_0=0.2$ and $\lambda_0=0.8$.

\subsubsection{X-ray temperature and luminosity functions}
In Section \ref{xrb}, we have emphasized the difference between our
formulation and the conventional PS approach. For more direct
comparison of these two procedures, we also calculate the X-ray
temperature and luminosity functions of clusters of galaxies.  

In our model, the X-ray temperature function at redshift $z$ is
expressed using equation (\ref{fdist}) as
%%%%%%%%%%%%%%%%%%%%%%%%%%%%%%%%%%%%%%%%%%%%%%%%%%%%%%%%%%%%%%%%%%%%%%%%
\begin{equation} 
N_{\rm T}(T,z) dT = \left. \int_{z}^{\infty} dz_{\rm f} 
     F(M,z_{\rm f},z) \frac{dM}{dT} \right|_{M=M(T,z_{\rm f})}dT, 
\label{tempfnf}
\end{equation}
%%%%%%%%%%%%%%%%%%%%%%%%%%%%%%%%%%%%%%%%%%%%%%%%%%%%%%%%%%%%%%%%%%%%%%%%
where $M(T,z_{\rm f})$ stands for the mass which gives the gas temperature
$T$ if formed at redshift $z_{\rm f}$ (e.g., eq.~[\ref{virtemp}]). Here we
have neglected the intrinsic evolution of gas temperature. On the
other hand, from the PS mass function, one obtains
%%%%%%%%%%%%%%%%%%%%%%%%%%%%%%%%%%%%%%%%%%%%%%%%%%%%%%%%%%%%%%%%%%%%%%%%
\begin{equation} 
N_{\rm T}(T,z) dT = N_{\rm PS}(M,z)
\left.\frac{dM}{dT}\right|_{M=M(T,z_{\rm f}=z)} dT.
\label{tempfnps}
\end{equation}
%%%%%%%%%%%%%%%%%%%%%%%%%%%%%%%%%%%%%%%%%%%%%%%%%%%%%%%%%%%%%%%%%%%%%%%%
Note that $z_{\rm f}$ is set to be equal to $z$ in the above equation. 

Similarly, the X-ray luminosity function at redshift $z$ in each
approach is
%%%%%%%%%%%%%%%%%%%%%%%%%%%%%%%%%%%%%%%%%%%%%%%%%%%%%%%%%%%%%%%%%%%%%%%%
\begin{equation} 
N_{\rm L}(L,z) dL = \left. \int_{z}^{\infty} dz_{\rm f} 
       F(M,z_{\rm f},z)  \frac{dM}{dL} \right|_{M=M(L,z_{\rm f},z)} dL, 
\label{lumfnf}
\end{equation}
%%%%%%%%%%%%%%%%%%%%%%%%%%%%%%%%%%%%%%%%%%%%%%%%%%%%%%%%%%%%%%%%%%%%%%%%
and 
%%%%%%%%%%%%%%%%%%%%%%%%%%%%%%%%%%%%%%%%%%%%%%%%%%%%%%%%%%%%%%%%%%%%%%%%
\begin{equation} 
N_{\rm L}(L,z) dL= N_{\rm PS}(M,z)
\left.\frac{dM}{dL}\right|_{M=M(L,z_{\rm f}=z,z)} dL, 
\label{lumfnps}
\end{equation}
%%%%%%%%%%%%%%%%%%%%%%%%%%%%%%%%%%%%%%%%%%%%%%%%%%%%%%%%%%%%%%%%%%%%%%%%
where $M(L,z_{\rm f},z)$ is solved from the integral of equation
(\ref{luminosity}) over frequency.

%%%%%%%%%%%%%%%%%%%%%%%%%%%%%%%%%%%%%%%%%%%%%%%%%%%%%%%%%%%%%%%%%%%%%
\subsection{Comparison with hydrodynamical simulation}
It is meaningful to check the validity and limitations of our modeling
through the comparison with other classes of approach. Fortunately a
series of three-dimensional hydrodynamic simulations in a CDM model
universe have been developed to study the evolution of X-ray clusters
(Cen et al. 1990; Cen \& Ostriker 1992, 1993, 1994; Kang et al. 1994;
Bryan et al. 1994). In this subsection, therefore, we compare the
results of our analytical approach with those of numerical
simulations.

%##################################################################
\begin{figure}
\begin{center}
   \leavevmode\psfig{figure=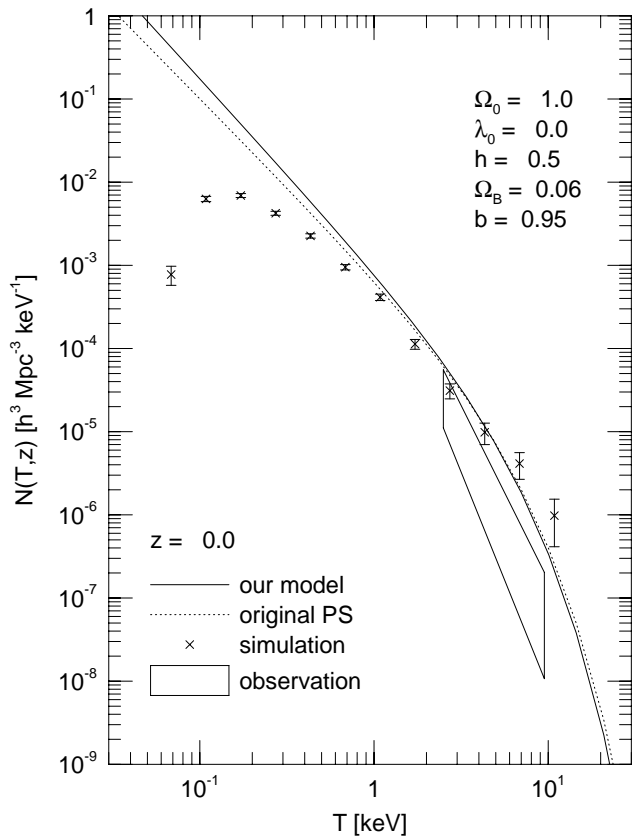,height=11cm}
\end{center}
\caption{X-ray cluster temperature functions at redshift $z=0$ in the
CDM model; predictions of our model
(eq.~[\protect\ref{tempfnf}\protect]) (solid curve), of the PS
approach (eq.~[\protect\ref{tempfnps}\protect]) (dotted curve), and of
the simulation by Kang et al. (1994) (crosses). The cosmological
parameters are fixed in each as $\Omega_0=1$, $h=0.5$, $\Omega_{\rm
B}=0.06$ and $b=0.95$.  The error bars of the simulation data only
indicate the statistical error. The squared region shows the error box
of observations (Henry \& Arnaud 1991). }
\label{figtemp}
\end{figure}
%##################################################################
Figure~4 illustrates the temperature functions evaluated analytically
by equations (\ref{tempfnf}) and (\ref{tempfnps}) respectively, in
comparison with the results of the simulation by Kang et al. (1994)
and the observations (Henry \& Arnaud 1991). The cosmological
parameters are the same as in Kang et al. (1994); $\Omega_0=1$,
$h=0.5$, $\Omega_{\rm B}=0.06$ and $b=0.95$. The indicated error bars
of the simulation data correspond to statistical ones only.  The
simulation data are kindly provided by Renyue Cen, and note that the
original plots in Figures 1 to 5 of Kang et al. (1994) should be
shifted down by a factor of $\ln 10$ due to an error in their
plotting procedure (Cen, private communication).

The analytical results in Fig.~4 show a reasonable agreement with that
of the numerical simulation above $T \sim 1 \rm{keV}$. In fact, the
simulation data are subject to much larger error below this
temperature due to its spatial resolution of $\sim 0.75 h^{-1}$Mpc
(Kang et al. 1994). They are also likely to be affected by the limited
sample volume size at $T\ngt 10$keV. The difference between our method
and the previous PS approach is in general rather insignificant in
this figure, because the temperature of a virialized system depends
only weakly on $z_{\rm f}$ as $\propto (1+z_{\rm f})$ in the
$\Omega_0=1$ universe (eq.~[\ref{virtemp}]). Such a feature is
especially apparent at high temperatures, where most clusters that
contribute are large and formed only at low redshifts. The difference
becomes relatively larger at low temperatures where small clusters
make dominant contributions.

%##################################################################
\begin{figure}
\begin{center}
   \leavevmode\psfig{figure=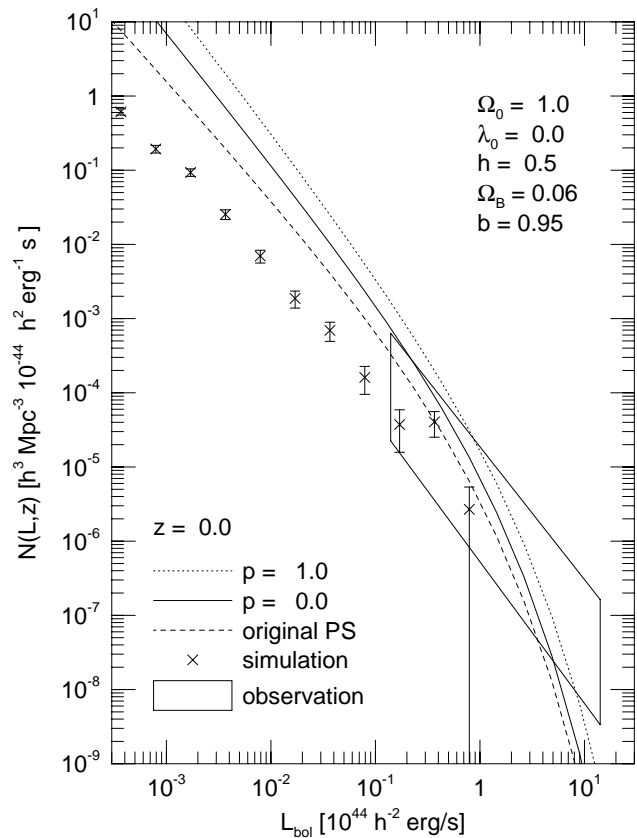,height=11cm}
\end{center}
\caption{X-ray cluster bolometric luminosity functions at redshift
$z=0$ in the CDM model; predictions of our model
(eq.~[\protect\ref{tempfnf}\protect]) with the parameter $p=0$ (solid
curve)and $p=1$ (dotted curve), of the PS formalism
(eq.~[\protect\ref{tempfnps}\protect]) (dashed curve), and of the
simulation by Kang et al. (1994) (crosses). The cosmological
parameters are the same as in Fig.~\protect\ref{figtemp}\protect. The
error bars of the simulation data are only statistical.  The squared
region shows the error box of observations (Henry \& Arnaud 1991). }
\end{figure}
%##################################################################
The results for the bolometric luminosity functions are plotted in
Fig.~5.  Neither the simulation nor our analytical approach takes
account of physical cooling or heating. This is justified in
discussing the overall properties of clusters of galaxies except at
the core. In contrast to Fig.~4, the difference between our method and
the conventional PS approach is more significant, because the
bolometric luminosity depends much more strongly on $z_{\rm f}$ as
$\propto (1+z_{\rm f})^{7/2}$ (eq.~[\ref{bollum}]). The deviations
from the simulation data are also larger and our changes to the PS
formalism does not reconcile the discrepancy. However, as remarked by
Cen (1992) and Kang et al. (1994), the numerical procedures tend to
systematically underestimate the predicted luminosity, perhaps by a
factor of $\sim 2$. The underestimation would be severer at low
luminosities where the effects of the spatial resolution becomes more
important. In addition, the simulation clearly lacks the sample volume
size for the statistically reliable predictions at $L_{\rm X} \ngt
10^{44} h^{-2}$erg/s where most of the observed data are available
(Henry \& Arnaud 1991). Thus the analytical results are not only
qualitatively consistent with the numerical simulation, but also
applicable in much wider dynamical range.

%##################################################################
\begin{figure}
\begin{center}
   \leavevmode\psfig{figure=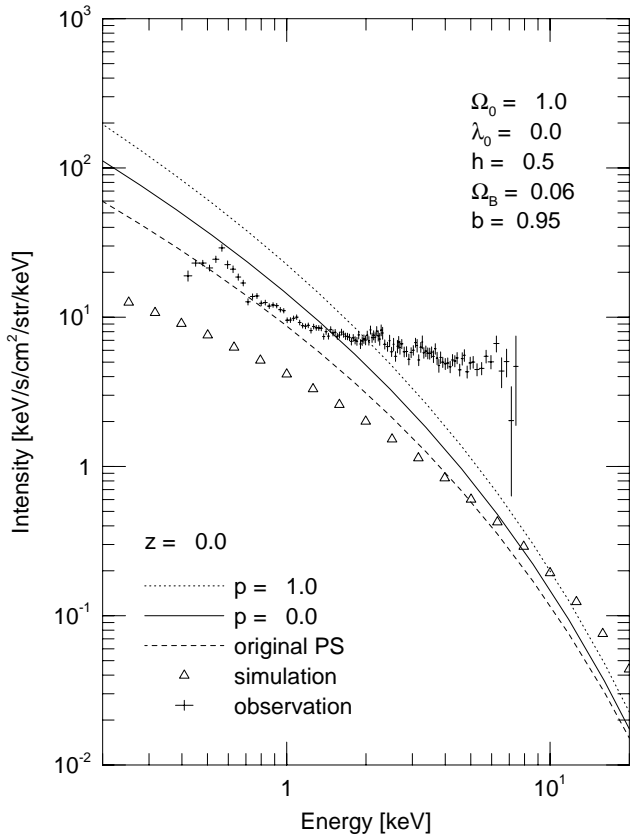,height=11cm}
\end{center}
\caption{The XRB spectra contributed from clusters of galaxies;
predictions of our model (eq.~[\protect\ref{power}\protect]) with the
parameter $p=0$ (solid curve) and $p=1$ (dotted curve), of the PS
formalism (eq.~[\protect\ref{power2}\protect]) (dashed curve), and of
the simulation by Kang et al. (1994) (triangles). The cosmological
parameters are the same as in Fig.~\protect\ref{figtemp}\protect. Also
plotted are the observed data from {\it ASCA} with statistical errors
(Gendreau et al. 1994). }
\end{figure}
%##################################################################
Figure~6 shows the predictions of the XRB spectra, together with the
{\it ASCA} observations. The analytical results agree well with the
simulation around $E\sim 10$ keV but predict larger intensity at $E
\nlt 2$ keV. This is reasonable if one considers the numerical effects
including the spatial resolution of the simulation. As expected from
the features in Fig.~5, our model systematically predicts higher
intensity than the original PS formalism, and the intensity increases
as the intrinsic evolution indicated by the parameter $p$ becomes
stronger.

As a matter of fact, our method and numerical simulations are
complementary to each other in predicting the X-ray properties of
clusters and their contribution to the XRB spectra. The analytical
approach basically has two advantages; one is that it is much easier
to make systematic parameter surveys. Therefore it would give a better
insight into how the predictions depend on the distribution of
intracluster gas, the luminosity evolution, and the geometry of the
background universe. The other is that the results are largely free
from the small-scale resolution or the limited volume size that may
affect the numerical results. Alternatively, our present predictions
do not properly take account of the thermal evolution of clusters
induced by the interaction between gas and radiation, which will be
incorporated in due course. In addition, we have assumed a rather
simplified gas density profile as well as a spherical distribution of
dark matter.

%%%%%%%%%%%%%%%%%%%%%%%%%%%%%%%%%%%%%%%%%%%%%%%%%%%%%%%%%%%%%%%%%%%%%%%%
\subsection{Predictions for various cosmological models}
\label{prediction}

In this subsection, we present the XRB spectra to constrain the viable
sets of cosmological parameters in a CDM universe. Specifically we
consider six CDM models listed in Table 1.  Values of the Hubble
parameter $h$ are taken as either $0.5$ or $0.8$, where the latter is
strongly motivated from the observation of Cepheids in the Virgo
cluster by the Hubble Space Telescope (Freedman et al. 1994).  In each
model, the baryon density of the universe is assumed to be
$\Omega_{\rm B} h^2=0.0125$, which is consistent with primordial
nucleosynthesis (e.g., Walker et al. 1991). The biasing parameter $b$
is given by normalizing the amplitude of the fluctuation spectrum by
the {\it COBE} DMR two year data (Sugiyama 1995).  The critical
overdensity for the spherical collapse $\delc$ is taken as
$\delc=1.69$ in models E5, E8, L5 and L8, \hspace{2mm} and
$\delc=1.64$ in O5 and O8 (Lilje 1992; LC).
%##############################################################
\begin{table}
\caption{Models for the background universe}
\begin{center}
\begin{tabular}{@{}c c c c c c}
Model &  $\Omega_0$ & $\lambda_0$ & $h$  &$\Omega_{\rm B}$& $b$  \\[1mm]
E5    &     1      &    0         &  0.5 &      0.05      & 0.71 \\
E8    &     1      &    0         &  0.8 &      0.02      & 0.43  \\
O5    &     0.2    &    0         &  0.5 &      0.05      & 6.7  \\
O8    &     0.2    &    0         &  0.8 &      0.02      & 2.8  \\
L5    &     0.2    &    0.8       &  0.5 &      0.05      & 2.2 \\
L8    &     0.2    &    0.8       &  0.8 &      0.02      & 1.0 
\end{tabular}
\end{center}
\end{table}
%#############################################################
%##################################################################
\begin{figure*}
\begin{center}
   \leavevmode\psfig{figure=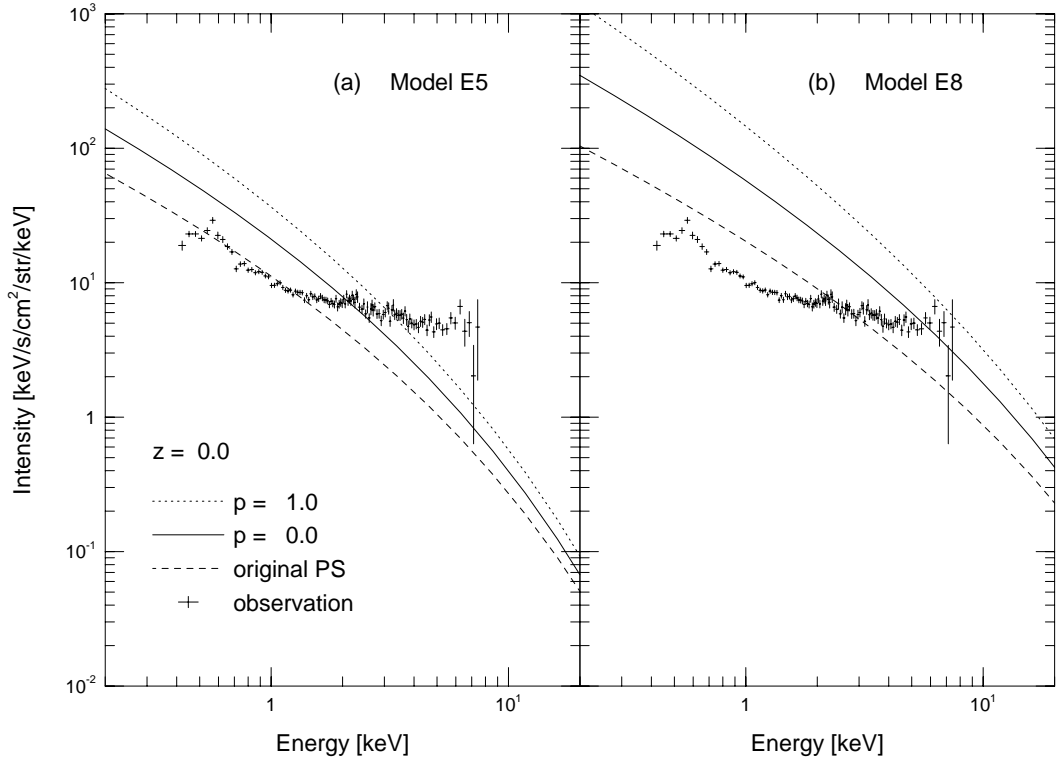,height=10cm,angle=90}
\end{center}
\caption{The XRB spectra contributed by clusters of galaxies in the
Einstein-de Sitter universe; (a) E5, (b) E8. Lines show predictions of
our model with the parameter $p=0$ (solid) and $p=1$ (dotted), and of
the original PS formalism (dashed). Also plotted are the {\it ASCA}
observations from Gendreau et al. (1994). }
\end{figure*}
%##################################################################
%##################################################################
\begin{figure*}
\begin{center}
   \leavevmode\psfig{figure=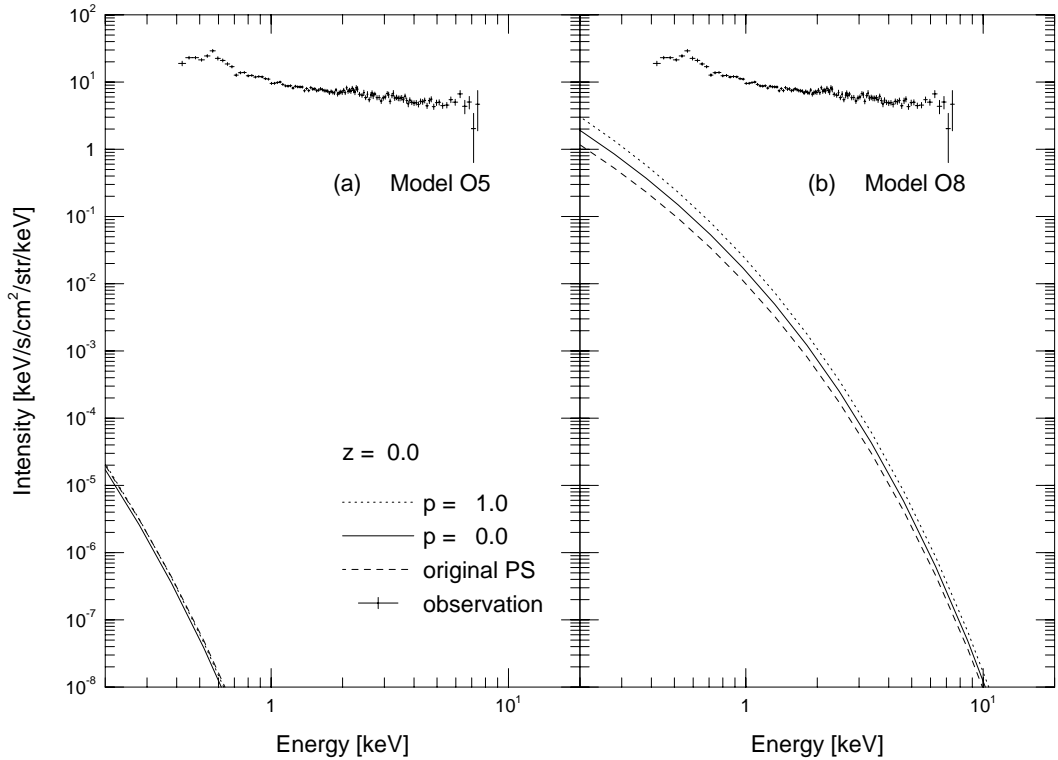,height=10cm,angle=90}
\end{center}
\caption{Same as in Fig.~7 except that the background universe is open
($\Omega_0=0.2$, $\lambda_0=0$); (a) O5, (b) O8. }
\end{figure*}
%##################################################################
%##################################################################
\begin{figure*}
\begin{center}
   \leavevmode\psfig{figure=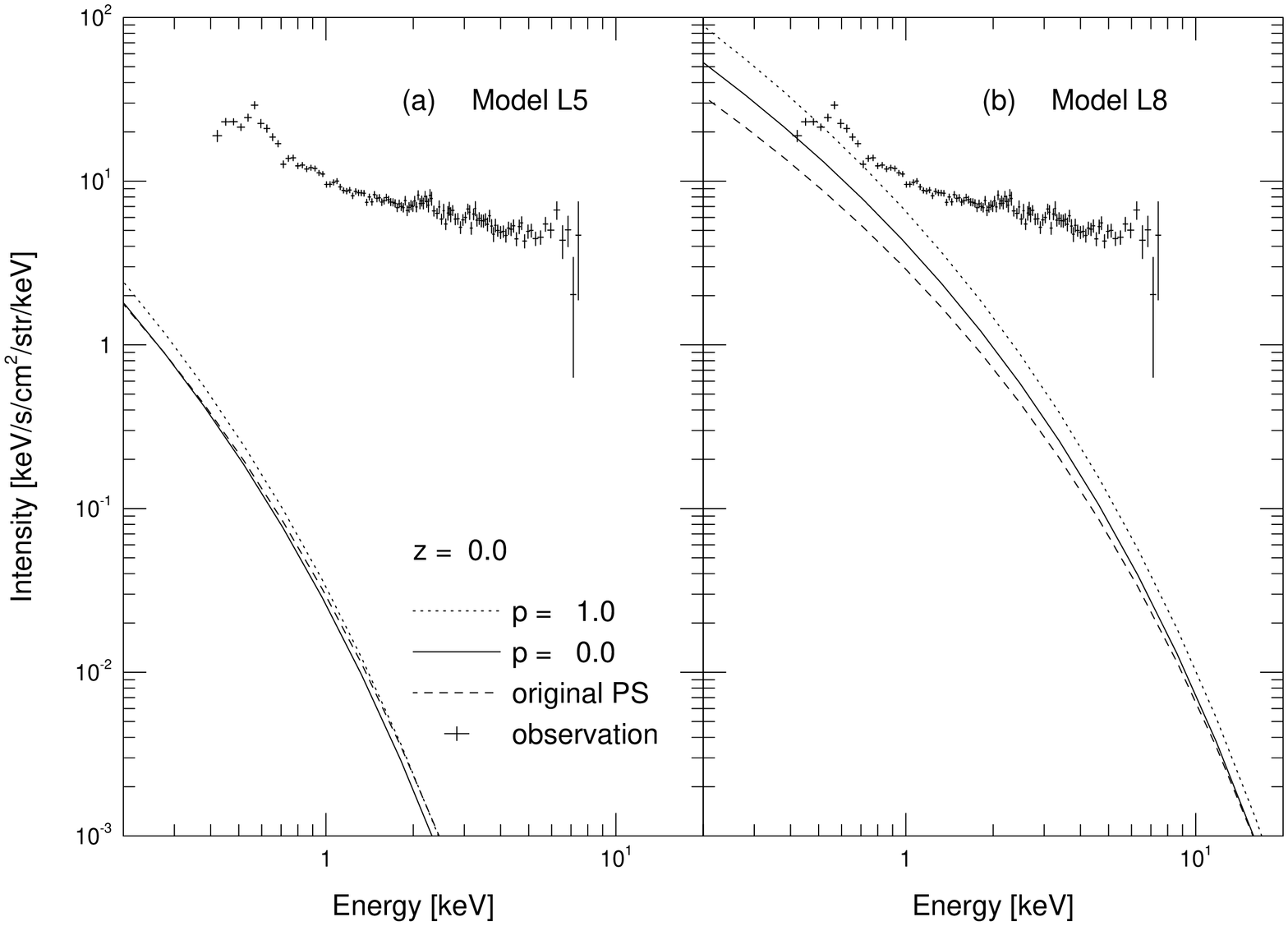,height=10cm,angle=90}
\end{center}
\caption{Same as in Fig.~7 except that the background universe is
spatially flat with the cosmological constant ($\Omega_0=0.2$,
$\lambda_0=0.8$); (a) L5, (b) L8.}
\end{figure*}
%##################################################################

Figures~7, 8 and 9 illustrate the calculated XRB spectra together with
the {\it ASCA} observations (Gendreau et al. 1994). There are a few
points that need to be kept in mind. Firstly, line emissions or
absorptions from metals are omitted in our calculation. This may lead
to underestimation of XRB intensity below 1 keV (Cen et al. 1995).
Secondly, a steep rise in the observed intensity in the soft X-ray
band may be partly due to Galactic emission (McCammon \& Sanders 1990;
Snowden et al. 1990).  Finally, it has been suggested from the
high-resolution image of {\it ROSAT} that at least $30 \%$ of the XRB
at 1 keV is contributed from quasars (Shanks et al. 1991).

Having considered the above points, we see that models E5 and E8
overproduce the soft X-ray intensity (Fig.~7). While this result was
pointed out earlier by Burg et al. (1993) on the basis of the
conventional PS approach, such a feature is more apparent in our model
which takes account of the formation rate.  The overproduction could be
removed if the clusters underwent considerable {\it anti-evolution}
against the growth of the gravitational potential (Burg et al. 1993).
Alternative ways to avoid such discrepancy are to consider a small
value for the Hubble constant; e.g., $h \nlt 0.3$, or to abandon the
{\it COBE} normalization and introduce much smaller fluctuation
amplitude, both of which are highly artificial even if not completely
excluded. Therefore the $\Omega_0=1$ CDM universes have a serious
difficulty of overproducing the XRB.

On the other hand, Fig.~8 shows that the predicted intensity in models
O5 and O8 is significantly smaller and that clusters of galaxies do
not basically contribute to the XRB in these models.  As a matter of
fact, this is mainly because of large values of the biasing parameter
$b$ imposed from the COBE data. Clusters of galaxies are thus rejected
as a candidate for the origin of the XRB in these open models.

Noticeably, Fig.~9 suggests that the observed XRB at $E \nlt 0.8$keV
attains a dominant contribution from clusters of galaxies in model L8. 
We should, however, be careful because the observation in this band is
most sensitive to the contamination due to Galactic absorptions and
emissions. Nevertheless it is interesting to see a possibility that a
part of the XRB can be naturally explained by clusters of galaxies in
one of promising cosmological models.

%%%%%%%%%%%%%%%%%%%%%%%%%%%%%%%%%%%%%%%%%%%%%%%%%%%%%%%%%%%%%%%%%%%%%%
%%%%%%%%%%%%%%%%%%%%%%%%%%%%%%%%%%%%%%%%%%%%%%%%%%%%%%%%%%%%%%%%%%%%%%
\section{CONCLUSIONS}

We have extended the theory of Press \& Schechter (1974) and the
formulation of Bower (1991) and Lacey \& Cole (1993) to derive
expressions for the rates of formation and destruction of
gravitationally bound systems.  Although the formal expressions of
these quantities diverge, they properly reproduce the time derivative
of the PS mass function. The divergence is not unphysical but can be
interpreted as being originated from the somewhat improper definition
of formation and destruction of bound objects.  We have proposed a
phenomenological prescription to remove such divergences by specifying
a realistic mass range in which the identity of an object is to be
maintained. The resulting expressions are practically insensitive to
the choice of the threshold masses. On the basis of this prescription,
we have obtained a distribution function of formation epochs of bound
systems.  This quantity is of great cosmological importance because it
provides a useful theoretical tool in describing the subsequent
evolution of the systems.

In order to exhibit the applicability of the above formalism, we have
evaluated the contribution of clusters of galaxies to the XRB, as well
as the X-ray cluster temperature and luminosity functions. Changes
from the previous PS formalism are obvious especially when the cluster
evolution is strong and the X-ray properties depend largely on its
formation epoch. The results of our analytical procedure, although
fairly phenomenological in modelling clusters' X-ray gas, are
reasonably supported by those of the three-dimensional hydrodynamical
simulation by Kang et al. (1994).  Thus these two approaches are
complementary in investigating the evolution history of virialized
structures.

We have shown that the significant fraction of the observed soft XRB
can be accounted for by clusters of galaxies in a spatially flat CDM
universe with $\Omega_0 \sim 0.2$ and $\lambda_0=1-\Omega_0$ with the
{\it COBE} DMR two year normalization. The $\Omega_0=1$ CDM model
tends to overproduce the observed X-ray intensity.  On the other hand,
the cluster contribution becomes negligibly small in an open CDM
universe with $\Omega_0 \sim 0.2$, in which case observed XRB should
be explained entirely by other classes of sources such as quasars.
Although our overall conclusion in the $\Omega_0=1$ case agrees with
Burg et al. (1993), we derived new constraints on the other models from
the XRB in a consistent and clear manner.  Note that the model with
$\Omega_0 \sim 0.2$, $\lambda_0=1-\Omega_0$ and $h\sim 0.8$ is a
specific example of the most successful cosmological models explaining
the cosmic age, two-point correlation functions of galaxies and
clusters, and small-scale velocity dispersions (Efstathiou et
al. 1990; Suginohara \& Suto 1991; Suto 1993; Watanabe, Matsubara \&
Suto 1994; Bahcall 1994). In this context it is interesting to note
that the same model can simultaneously account for a major fraction of
the observed XRB in the soft energy band.  Furthermore, the microwave
background distortion and the resulting anisotropy due to the distant
clusters in CDM models are below the current limit (Makino \& Suto
1993).

Obviously there remain several areas for further theoretical
work. With regard to the XRB predictions, the greatest uncertainty
probably lies in the evolution of intracluster medium. Although we
have assumed that it basically traces the growth of the gravitational
potential, this assumption is still an open question and different
evolutionary models would alter the predicted luminosity functions and
the XRB spectra. The contamination of the observed spectra by the
Galactic emission and absorption is also uncertain. Moreover, we have
omitted the line emissions which may affect the soft band intensity.
Therefore we did not aim to rigorously reproduce the XRB spectra from
theory, but rather to examine the possibility that clusters of
galaxies play a crucial role in the origin of the XRB.

As to the formalism, on the other hand, our method contains a formal
divergence for which somewhat artificial prescription has been
proposed. In addition, our argument itself has further room for
improvement; for instance, if $M_1$ in equation (\ref{formdef}) does
not correspond to the largest parent of the final object $M$, the rate
of formation thereby defined possibly overcounts the true rate.  This
problem is in fact ascribed to a fundamental limitation of the current
formalism which does not take into account the history of every single
object involved in a merger process.  Clearly the validity of our
treatment needs to be examined through a direct comparison with other
classes of approach such as N-body simulations. Other ways of
formulation are also desirable. In this context we note that LC
proposed a differential distribution function of halo formation times
based on what they called `the halo counting argument' (see their \S
2.5.2). As notified by LC, however, their definition leads to a slight
self-inconsistency of predicting a negative probability density
depending on the slope of the fluctuation spectrum.  As a matter of
fact, we compared their result with ours and basically confirmed that
both agree well in the mass range of astrophysical interest.  For
practical purposes, therefore, our present formulae are expected to be
applicable to a number of further problems in cosmology such as
cluster luminosity functions in various cosmological models, formation
of high redshift objects and reionization history of the
universe. These issues, together with a quantitative comparison with
LC, will be described elsewhere (Kitayama \& Suto, in preparation).

\vspace{1cm}
\noindent
{\large\bf ACKNOWLEDGEMENTS}\\[4mm]
\noindent
We thank Shin Sasaki, Katsuhiko Sato, Tatsushi Suginohara, Hajime
Susa, and Takahiro Tanaka for helpful discussions and comments. We are
grateful to Renyue Cen, Keith Gendreau and Naoshi Sugiyama for
providing the relevant results of their current work which are adopted
in the present paper. TK acknowledges the fellowship from the Nippon
Telephone and Telegram Co. .  This research is supported in part by
the Grants-in-Aid by the Ministry of Education, Science and Culture of
Japan (05640312, 06233209, 07740183, 07CE2002).

\appendix
\section{RADIUS OF VIRIALIZED CLUSTERS IN LOW $\Omega_0$ UNIVERSES}

In an open universe ($\Omega_0 < 1$, $\lambda_0 =0$), the effective
radius that a spherical perturbation attains when it virializes is
estimated as half of its maximum radius at turn-around.  If a
virialized system is formed with mass $M$ at redshift $z_{\rm f}$, one
can readily show that the radius $r_{\rm vir}$ is given by
%%%%%%%%%%%%%%%%%%%%%%%%%%%%%%%%%%%%%%%%%%%%%%%%%%%%%%%%%%%%%%%%%%%%%%%%
\begin{equation} 
r_{\rm vir}=\frac{(GM)^{1/3}}{1-\Omega_0} \mbkt{\frac{\Omega_0(\sinh
\eta_{\rm f}-\eta_{\rm f})}{4 \pi H_0}}^{2/3}
\label{ropen}
\end{equation}
%%%%%%%%%%%%%%%%%%%%%%%%%%%%%%%%%%%%%%%%%%%%%%%%%%%%%%%%%%%%%%%%%%%%%%%%
where $\eta_{\rm f}$ is expressed in terms of the epoch of collapse 
$z_{\rm f}$ as  
%%%%%%%%%%%%%%%%%%%%%%%%%%%%%%%%%%%%%%%%%%%%%%%%%%%%%%%%%%%%%%%%%%%%%%%%
\begin{equation}
\eta_{\rm f} = \mbox{arccosh}\lbkt{\frac{2(1-\Omega_0)}{\Omega_0(1+z_{\rm
f})} + 1}.
\end{equation}
%%%%%%%%%%%%%%%%%%%%%%%%%%%%%%%%%%%%%%%%%%%%%%%%%%%%%%%%%%%%%%%%%%%%%%%%

In the presence of the cosmological constant $\lambda_0$, the
dynamical motion of a spherical perturbation is modified and the
radius $r_{\rm vir}$ of a virialized cluster is no longer half of the
turn-around radius $r_{\rm ta}$.  In fact, the ratio of these two
quantities is shown to obey the following cubic equation (Lahav et al. 
1991):
%%%%%%%%%%%%%%%%%%%%%%%%%%%%%%%%%%%%%%%%%%%%%%%%%%%%%%%%%%%%%%%%%%%%%%%%
\begin{equation}
2 \chi \sbkt{\frac{r_{\rm vir}}{r_{\rm ta}}}^3
-(2+\chi)\sbkt{\frac{r_{\rm vir}}{r_{\rm ta}}} +1=0,
\label{cubic}
\end{equation}
%%%%%%%%%%%%%%%%%%%%%%%%%%%%%%%%%%%%%%%%%%%%%%%%%%%%%%%%%%%%%%%%%%%%%%%%
where
%%%%%%%%%%%%%%%%%%%%%%%%%%%%%%%%%%%%%%%%%%%%%%%%%%%%%%%%%%%%%%%%%%%%%%%%
\begin{equation}
\chi\equiv\frac{\lambda_0 H_0^2}{GM}r_{\rm ta}^3.
\label{chi}
\end{equation}
%%%%%%%%%%%%%%%%%%%%%%%%%%%%%%%%%%%%%%%%%%%%%%%%%%%%%%%%%%%%%%%%%%%%%%%%
The condition for a shell to turn around is $\chi<1$. Lahav et
al. (1991) found a following approximated solution to equation
(\ref{cubic}):
%%%%%%%%%%%%%%%%%%%%%%%%%%%%%%%%%%%%%%%%%%%%%%%%%%%%%%%%%%%%%%%%%%%%%%%%
\begin{equation}
\frac{r_{\rm vir}}{r_{\rm ta}}\approx \frac{2-\chi}{4-\chi}.
\label{lahav}
\end{equation}
%%%%%%%%%%%%%%%%%%%%%%%%%%%%%%%%%%%%%%%%%%%%%%%%%%%%%%%%%%%%%%%%%%%%%%%%
For $\lambda_0=0$, the above result reproduces $r_{\rm vir}/r_{\rm
ta}=1/2$.

Now the maximum turn-around radius $r_{\rm ta}$ is no longer expressed
as a simple analytical form if $\lambda_0$ is nonzero. For our present
purpose, however, we simply assume that the presence of $\lambda_0$
affects the dynamics of the perturbation only through equation
(\ref{cubic}) and it does not alter the value of $r_{\rm ta}$
itself. Then in a spatially flat universe ($\Omega_0 < 1$, $\lambda=
1- \Omega_0$) which we discussed in the present paper, $r_{\rm ta}$ is
written down as
%%%%%%%%%%%%%%%%%%%%%%%%%%%%%%%%%%%%%%%%%%%%%%%%%%%%%%%%%%%%%%%%%%%
\begin{equation} r_{\rm ta}= 2(GM)^{1/3} \sbkt{\frac{\zeta_{\rm
f}}{3\pi H_0 \sqrt{1-\Omega_0}}}^{2/3},
\label{maxrad}
\end{equation}
%%%%%%%%%%%%%%%%%%%%%%%%%%%%%%%%%%%%%%%%%%%%%%%%%%%%%%%%%%%%%%%%%%%
where $\zeta_{\rm f}$ is related to $z_{\rm f}$ via  
%%%%%%%%%%%%%%%%%%%%%%%%%%%%%%%%%%%%%%%%%%%%%%%%%%%%%%%%%%%%%%%%%%%
\begin{equation}
\zeta_{\rm f}\equiv \mbox{arcsinh}\lbkt{
\sqrt{\sbkt{\frac{1}{\Omega_0}-1}\frac{1}{(1+z_{\rm f})^3}}\; }.
\label{zeta}
\end{equation}
%%%%%%%%%%%%%%%%%%%%%%%%%%%%%%%%%%%%%%%%%%%%%%%%%%%%%%%%%%%%%%%%%%%
\end{document}